\documentclass[twocolumn]{revtex4}
\usepackage{amssymb}
\usepackage{latexsym}
\usepackage{epsfig}

\usepackage{amsmath}
\begin{document}
\title{Note on Tsallis Holographic Dark Energy}
\author{M. Abdollahi Zadeh$^{1}$\footnote{m.abdollahizadeh@shirazu.ac.ir}, A.
Sheykhi$^{1,2}$\footnote{Corresponding
author:asheykhi@shirazu.ac.ir}, H. Moradpour $^{2}$\footnote{
h.moradpour@riaam.ac.ir} and Kazuharu
Bamba$^3$\footnote{bamba@sss.fukushima-u.ac.jp}}
\address{$^1$ Physics Department and Biruni Observatory, College of
Sciences, Shiraz University, Shiraz 71454, Iran\\
$^2$ Research Institute for Astronomy and Astrophysics of Maragha
(RIAAM), P.O. Box 55134-441, Maragha, Iran\\
$^3$ Division of Human Support System, Faculty of Symbiotic
Systems Science, Fukushima University, Fukushima 960-1296, Japan}

\begin{abstract}
We explore the effects of considering various infrared (IR)
cutoffs, including the particle horizon, the Ricci
horizon and the Granda-Oliveros (GO) cutoffs, on the
properties of Tsallis holographic dark energy (THDE) model,
proposed inspired by Tsallis generalized entropy formalism
\cite{THDE}. Interestingly enough, we find that for the particle
horizon as IR cutoff, the obtained THDE model can describe the
late time accelerated universe. This is in contrast to the usual
HDE model which cannot lead to an accelerated universe, if one
considers the particle horizon as the IR cutoff.
We also investigate the cosmological consequences of THDE under
the assumption of a mutual interaction between the dark sectors of
the Universe. It is shown that the evolution history of the
Universe can be described by these IR cutoffs and thus the current
cosmic acceleration can also be realized. The sound
instability of the THDE models for each cutoff are also
explored, separately.
\end{abstract}
\maketitle

\section{Introduction}
There are various cosmological observations which indicate that
our Universe is now experiencing an accelerated expansion phase
\cite{Riess,Riess1,Riess2,Riess3,COL2001,COL20011,COL20012,COL20013,
Ade2014,Astier,Astier2006,Astier20061,Astier20062,Astier20063,Astier20064}.
The origin of this accelerated phase is attributed to a
mysterious matter which is called dark energy (DE). For reviews on
the DE problem and the modified gravity theories, which is called
geometric DE, to account for the late-time cosmic acceleration,
see, e.g.,~\cite{R-DE-MG}. In this regard, the holographic dark
energy (HDE) is an interesting attempt which can address
this bizarre problem in the framework of quantum gravity by using
the holographic hypothesis \cite{Hooft,Hooft1995,Cohen}. This
model is in agreement with various astronomical observations
\cite{Zhang2005,Zhang20051,Zhang20052,Zhang20053,Zhang20054}, and
various scenarios of it can be found in
\cite{Li2004,Hsu,Nojiri:2005pu, GO1, GO2, SSB,Wang1,
Wang2,cop,Zim,shey0,shey1,shey2}. Horizon entropy is the backbone
of the HDE models, and hence, any change of the horizon entropy
affects the HDE model. Another important player in these models is
the IR cutoff, and indeed, the various IR cutoffs lead to
different HDE models \cite{Li2004,GO1, GO2}.

Since gravity is a long-range interaction, one can also use the
generalized statistical mechanics to study the gravitational
systems \cite{Majhi,Abe,Sayahian,Renyi,Komatsu,Tsallis}. In this
regard, due to the fact that the black hole entropy can be
obtained by applying the Tsallis statistics to the system
\cite{Majhi,Abe}, three new HDE models with titles THDE, SMHDE and
RHDE have recently been proposed \cite{THDE,Sayahian,Renyi}. Among
these three models, in the absence of an interaction between the
cosmos sectors, RHDE, based on the Renyi entropy and the first law
of thermodynamics, shows more stability by itself
\cite{Renyi}. In fact, in a noninteracting universe, while SMHDE
is classically stable whenever SMHDE is dominant in the universe,
THDE, built using the Tsallis generalized entropy \cite{Tsallis},
is never stable at the classical level \cite{THDE,Sayahian}. It is
also worth mentioning that a THDE model whose IR cutoff is the
future event horizon has been studied in a noninteracting universe
showing satisfactory results \cite{bam}.

On the other side, the cosmological observations admit an
interaction between the two dark sectors of cosmos including DE
and DM \cite{gdo1,gdo2,ob1,ob2,ob3,ob4,ob5,ob6,ob7,pav,h1}. The
existence of such mutual interaction may provide a solution for
the coincidence problem \cite{ob7,co1,co2,co3,co4,co5,co6,pav,h1}.
In the present work, we are interested in studying the
dynamics of a flat FRW universe filled with a pressureless source
and THDE in both interacting and non-interacting cases. In order
to build  THDE, we shall employ various IR cutoffs, including the
apparent and the particle horizons together with the GO
and the Ricci cutoffs.

The organization of this paper is as follows. In the next section,
we study the evolution of the Universe by considering an
interaction between DM and THDE whose IR cutoff is the apparent
horizon. Thereinafter, a new THDE is built by employing the
particle horizon as the IR cutoff, and then, the cosmic evolution
are investigated for both interacting and non-interacting
universes in section III. The cases of the GO and Ricci cutoffs
are studied in Secs. IV and V, respectively. The last section is
devoted to a summary and concluding remarks.

%%%%%%%%%%%%%%%%%%%%%%%%%%%%%%%%%%%%%%%%%%%%%%%%%%%%%%%%%%%%%%%%%%%%%%%%%%%%%%%%%%%%%%%%%%%%%%%
\section{Interacting THDE with Hubble Cutoff}\label{Interacting Hubble}
The energy density of THDE model is given by \cite{Tsallis}
\begin{eqnarray}\label{Trho}
\rho_D=BL^{2\delta-4},
\end{eqnarray}
where $B$ is an unknown parameter. We consider a
homogeneous and isotropic flat Friedmann-Robertson-Walker (FRW)
universe which is described by the line element
\begin{eqnarray}\label{frw}
ds^{2}=-dt^{2}+a^{2}\left( t\right) \left[ dr^2
+r^{2}d\Omega^{2}\right],
\end{eqnarray}
where $a(t)$ is the scale factor. The first Friedmann equation
takes the form
\begin{eqnarray}\label{frd}
H^{2}=\frac{1}{3m_{p}^{2}}\left(\rho_{D}+\rho_{m}\right),
\end{eqnarray}
where, $\rho_m$ and $\rho_D$ denote the energy density of
dark matter (DM) and THDE, respectively. Defining, as usual, the
dimensionless density parameters as
\begin{eqnarray}\label{3}
\Omega_{D}=\frac{\rho_{D}}{\rho_c}=\frac{B}{3m_{p}^{2}}H^{-2\delta+2}, \ \
\Omega_{m}=\frac{\rho_{m}}{\rho_c},
\end{eqnarray}
where $\rho_c=3m_{p}^{2}H^{2}$ is called the critical energy
density, we can easily rewrite the first Friedmann equation in the
form
\begin{eqnarray}\label{u}
\Omega_{m}+\Omega_{D}=1.
\end{eqnarray}
Moreover, we assume that DM and DE interact with each other
meaning that the conservation law is decomposed as
\begin{eqnarray}\label{conm}
&&\dot{\rho}_m+3H\rho_m=Q,\\
&&\dot{\rho}_D+3H(1+\omega_D)\rho_D=-Q,\label{conD}
\end{eqnarray}
in which $\omega_D\equiv{p_D}/{\rho_D}$ is the equation of state
(EoS) parameter of THDE and $Q$ denotes the interaction term
between DE and DM. Throughout this paper, $Q=3b^2
H(\rho_m+\rho_D)$, where $b^2$ is a coupling constant, is
considered as the mutual interaction between the cosmos sectors
\cite{pav,h1}. The ratio of the energy densities is also evaluated
as
\begin{equation}\label{r2}
r=\frac{\Omega_m}{\Omega_D}=\frac{1-\Omega_D}{\Omega_D}.
\end{equation}
Taking the time derivative of Eq.(\ref{frd}), and by using
Eqs.(\ref{conm}), (\ref{conD}) and (\ref{r2}), we can obtain
\begin{eqnarray}\label{7}
\frac{\dot{H}}{H^{2}}=-\frac{3}{2}(1+\omega_{D}+r)\Omega_{D}.
\end{eqnarray}
In addition, by considering the Hubble horizon as the IR cutoff,
$L=H^{-1}$, the energy density \ref{Trho} takes form
\begin{eqnarray}\label{Hrho}
\rho_D=BH^{-2\delta+4},
\end{eqnarray}
The time derivative of above equation, combined with
Eqs.(\ref{conD}) and~(\ref{7}), also leads to
\begin{eqnarray}\label{w1}
\omega_{D}=\dfrac{\delta-1+\frac{b^2}{\Omega_D}}{(2-\delta)\Omega_{D}-1}.
\end{eqnarray}
Simple calculations for the deceleration parameter, defined as
\begin{eqnarray}\label{q}
q=-1-\frac{\dot{H}}{H^{2}},
\end{eqnarray}
yield
\begin{eqnarray}\label{q1}
q=\left[\dfrac{(1-2\delta)\Omega_{D}+1-3b^2}{2(1-(2-\delta)\Omega_{D})}\right],
\end{eqnarray}
where we used Eqs.(\ref{w1}) and~(\ref{7}) to obtain this result.
Combining the time derivative of Eq. (\ref{3}) with Eqs. (\ref{7})
and (\ref{w1}), and defining $\Omega_{D}^{\prime}={d\Omega_D}/{d
(\ln a)}$, we get
\begin{eqnarray}\label{Omega}
\Omega_{D}^{\prime}=3(\delta-1)\Omega_{D}
\left(\dfrac{1-\Omega_{D}-b^2}{1-(2-\delta)\Omega_{D}}\right),
\end{eqnarray}
for the $L=H^{-1}$ case.

\begin{figure}[htp]
\begin{center}
\includegraphics[width=8cm]{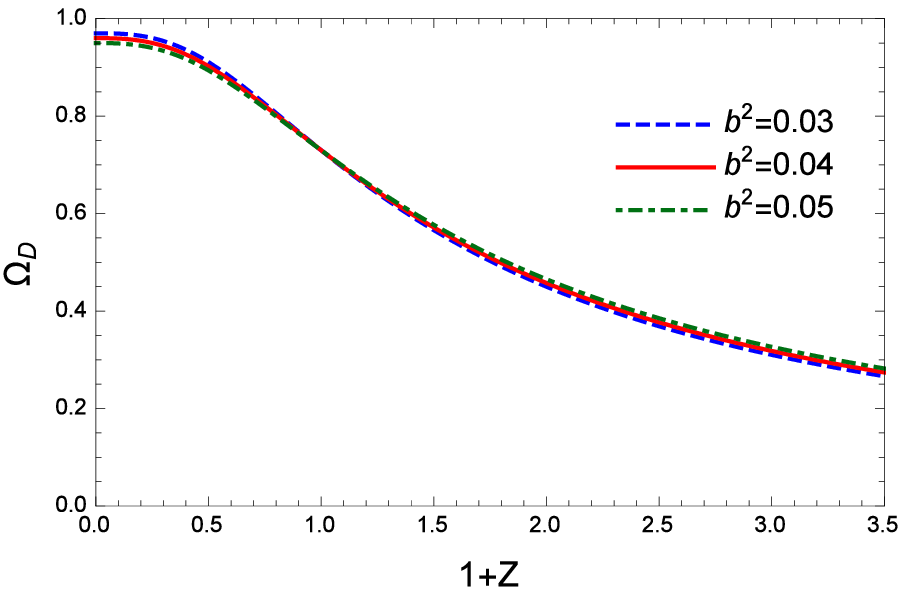}
\caption{The evolution of $\Omega_D$ versus redshift
parameter $z$ for
 interacting THDE with Hubble horizon as the
IR cutoff. Here, we have taken $\Omega^{0}_D=0\cdot73$ and
$\delta=1\cdot4$.}\label{Omega-z1}
\end{center}
\end{figure}
\begin{figure}[htp]
\begin{center}
\includegraphics[width=8cm]{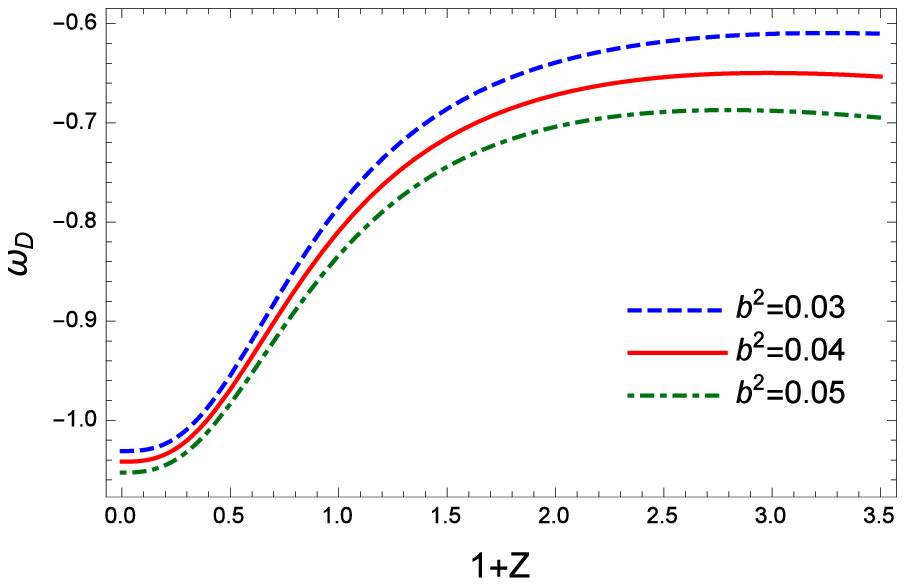}
\caption{The evolution of $\omega_D$ versus redshift
parameter $z$ for interacting THDE with Hubble horizon as
the IR cutoff. Here, we have taken $\Omega^{0}_D=0\cdot73$
and $\delta=1\cdot4$.}\label{w-z1}
\end{center}
\end{figure}
\begin{figure}[htp]
\begin{center}
\includegraphics[width=8cm]{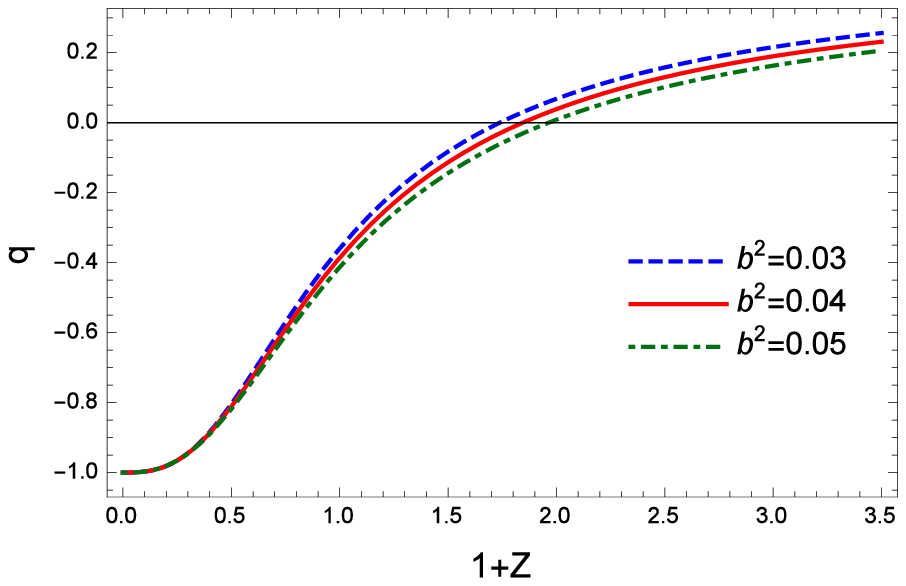}
\caption{The evolution of $q$ versus redshift parameter $z$
for interacting THDE with Hubble horizon as the IR cutoff.
Here, we have taken $\Omega^{0}_D=0\cdot73$ and
$\delta=1\cdot4$.}\label{q-z1}
\end{center}
\end{figure}

\noindent For $\delta=1\cdot4$ case and the initial condition
$\Omega^{0}_D=\Omega_D(z=0)=0.73$, the evolutions of $\Omega_D$,
$\omega_D$ and $q$ versus $(1+z)$ have been plotted in
Figs.~\ref{Omega-z1},~\ref{w-z1} and~\ref{q-z1}. From these
figures, one can see that $\omega_D$ can cross the phantom line,
and moreover, the value of the transition redshift is increased as
a function of $b^2$.
Finally, we explore the stability of the THDE model as
\begin{equation}\label{vs}
v_{s}^{2}=\frac{dP_D}{d\rho_D}=\frac{\dot{P}_D}{\dot{\rho}_D}=\dfrac{\rho_{D}}{\dot{\rho}_{D}}
\dot{\omega}_{D}+\omega_{D},
\end{equation}
Combining time derivative of Eq.(\ref{Hrho}) with Eq.~(\ref{7}), we have
\begin{eqnarray}\label{dotrho}
\dot{\rho}_{D}=3\rho_D(\delta-2)H(1+\Omega_D \omega_D),
\end{eqnarray}
which finally leads to
\begin{eqnarray}\label{vs1}
v_{s}^{2}=\dfrac{(\delta-1)(\Omega_{D}-1)+b^2[\delta+\frac{1}{(\delta-2)\Omega_D}]}{\left[1-(2-\delta)\Omega_{D}\right]^{2}},
\end{eqnarray}
where Eq. (\ref{vs}) and the time derivative of Eq. (\ref{w1}) have
been employed to obtain the above result. It is also useful
to note here that in the absence of interaction term $(b^2=0)$, Eqs.
(\ref{w1}), (\ref{q1}), (\ref{Omega}) and (\ref{vs1}) are reduced to
relations obtained in Ref \cite{THDE}.
\begin{figure}[htp]
\begin{center}
\includegraphics[width=8cm]{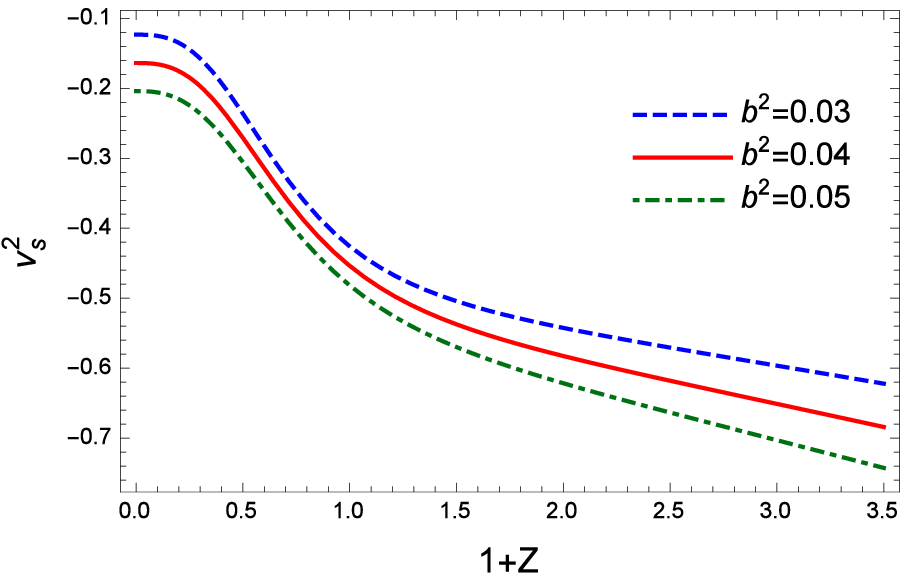}
\caption{The evolution of ${v}^{2}_{s}$ versus redshift
parameter $z$ for interacting THDE with Hubble horizon as
the IR cutoff. Here, we have taken $\Omega^{0}_D=0\cdot73$
and $\delta=1\cdot4$.}\label{vs-z1}
\end{center}
\end{figure}
\begin{figure}[htp]
\begin{center}
\includegraphics[width=8cm]{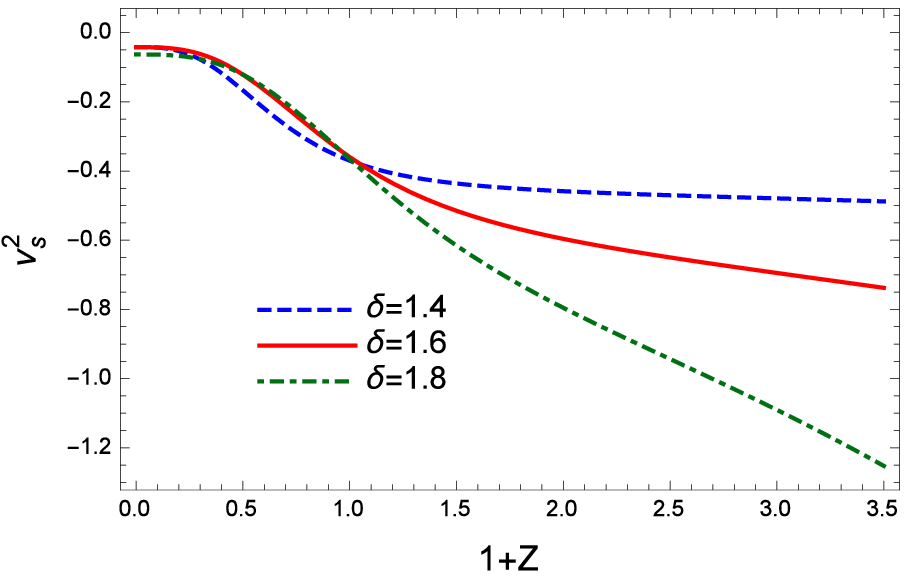}
\caption{The evolution of ${v}^{2}_{s}$ versus redshift
parameter $z$ for interacting THDE with Hubble horizon as
the IR cutoff. Here, we have taken $\Omega^{0}_D=0\cdot73$
and $b^2=0\cdot1$}\label{vs-z2}
\end{center}
\end{figure}
Figs.~\ref{vs-z1} and \ref{vs-z2} show that the interacting THDE
with Hubble cutoff is stable neither for a fixed $\delta$ nor for
a fixed $b^2$ meaning that the model is unstable, a result the
same as that of the non-interacting case \cite{THDE}.
%%%%%%%%%%%%%%%%%%%%%%%%%%%%%%%%%%%%%%%%%%%%%%%%%%%%%%%%%%%%%%%%%%%%%5
\section{THDE with particle horizon cutoff}\label{Interacting particle}
\subsection{Non-interacting}
\begin{figure}[htp]
\begin{center}
\includegraphics[width=8cm]{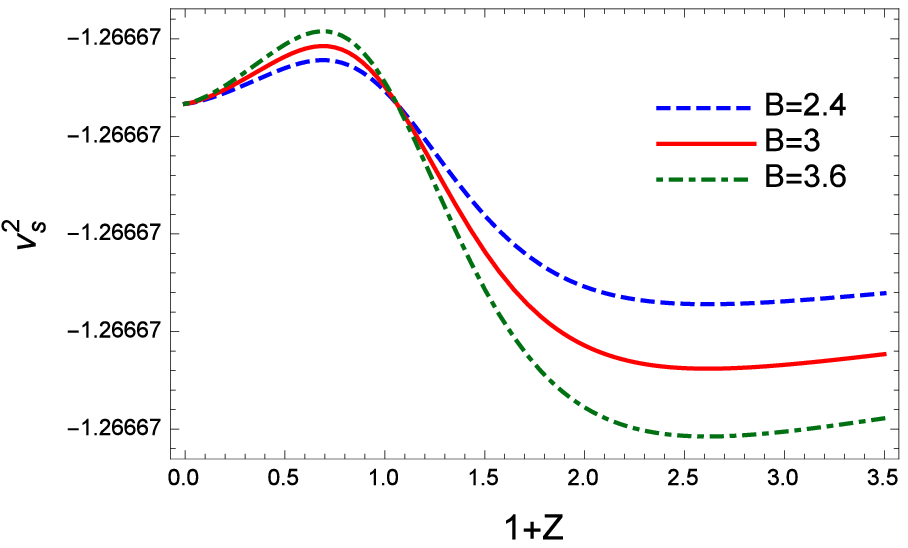}
\caption{The evolution of ${v}^{2}_{s}$ versus redshift
parameter $z$ for non-interacting THDE with particle horizon as
the IR cutoff. Here, we have taken $\Omega^{0}_D=0\cdot73$,
$\delta=2\cdot4$ and $H(a=1)=67$.}\label{Pvs-z3}
\end{center}
\end{figure}

\begin{figure}[htp]
\begin{center}
\includegraphics[width=8cm]{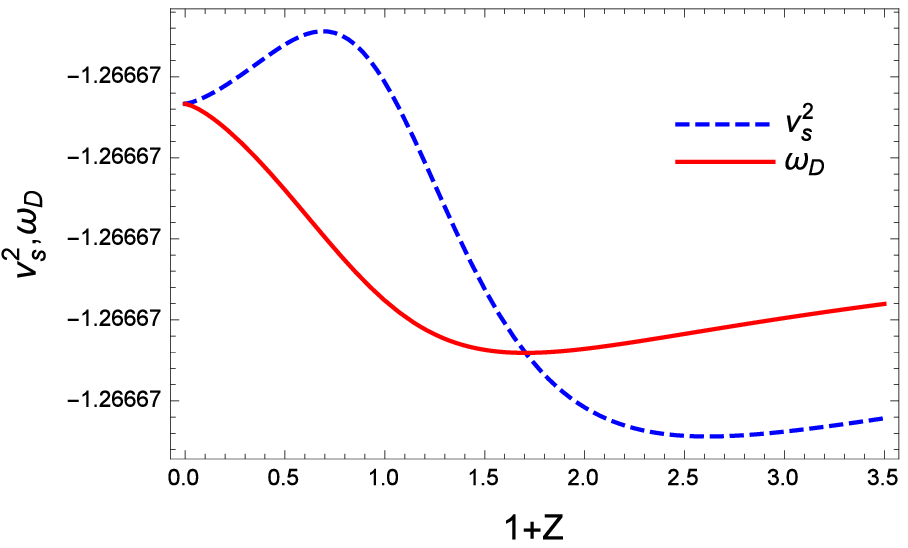}
\caption{The evolution of ${v}^{2}_{s}$ and $\omega_D$
versus redshift parameter $z$ for non-interacting THDE with particle
horizon as the IR cutoff. Here, we have taken
$\Omega^{0}_D=0\cdot73$, $B=2\cdot4$, $\delta=2\cdot4$ and
$H(a=1)=67$.}\label{Pvw1-z4}
\end{center}
\end{figure}

\begin{figure}[htp]
\begin{center}
\includegraphics[width=8cm]{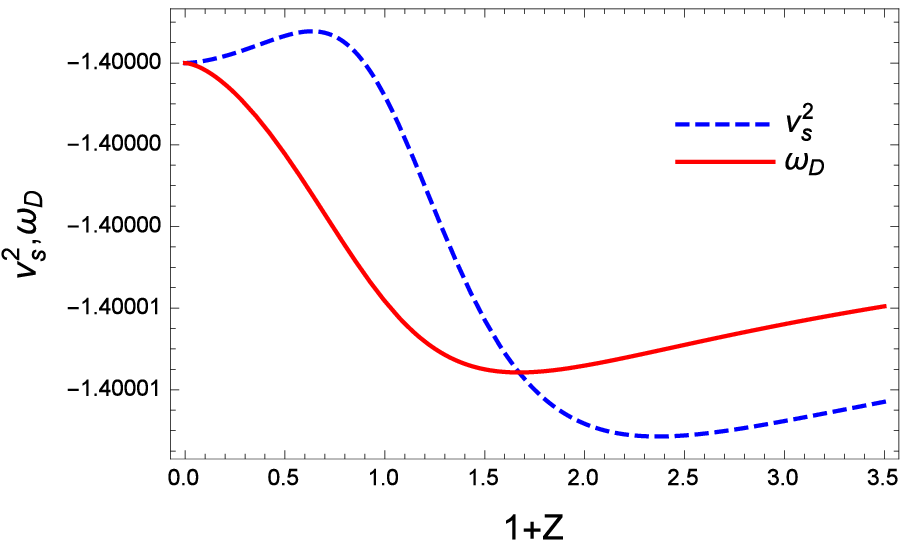}
\caption{The evolution of ${v}^{2}_{s}$ and $\omega_D$
versus redshift parameter $z$ for non-interacting THDE with particle
horizon as the IR cutoff. Here, we have taken
$\Omega^{0}_D=0\cdot73$, $B=2\cdot4$, $\delta=2\cdot6$ and
$H(a=1)=67$.}\label{Pvw2-z4}
\end{center}
\end{figure}

\begin{figure}[htp]
\begin{center}
\includegraphics[width=8cm]{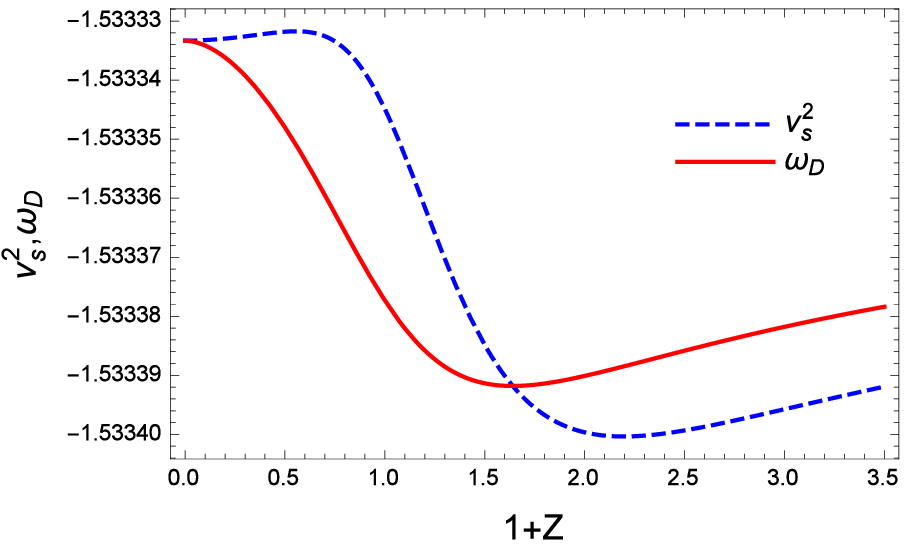}
\caption{The evolution of ${v}^{2}_{s}$ and $\omega_D$
versus redshift parameter $z$ for non-interacting THDE with particle
horizon as the IR cutoff. Here, we have taken
$\Omega^{0}_D=0\cdot73$, $B=2\cdot4$, $\delta=2\cdot8$ and
$H(a=1)=67$.}\label{Pvw3-z4}
\end{center}
\end{figure}

It is well-known that HDE model with particle horizon as
the IR cutoff cannot lead to an accelerated universe and it
is impossible to obtain an accelerated expansion \cite{Li2004}.
Indeed, with this cutoff, one always arrives at $\omega_D>-1/3$,
which is in contradiction with recent cosmological observations
\cite{Li2004}. As we shall see in this section, for the THDE with
particle horizon as the IR cutoff, it is quite possible to
reproduce an accelerating universe which is one of the main
advantages of THDE in comparison with the usual HDE model.
The particle horizon is defined as \cite{Li2004}
\begin{equation}
R_{p}=a(t)\int_{0}^{t}{\frac{dt}{a(t)}},
\end{equation}
which satisfies the following condition
\begin{equation}\label{particle}
\dot{R}_{p}=H R_p +1.
\end{equation}
Therefore, bearing Eq. (\ref{Trho}) in mind, the energy density of THDE is obtained as
\begin{eqnarray}\label{Prho}
\rho_D=B R_{p}^{2\delta-4},
\end{eqnarray}
where its time derivative leads to
\begin{eqnarray}\label{dotPrho}
\dot{\rho}_{D}=\rho_D(2\delta-4)H(1+F),
\end{eqnarray}
in which
\begin{eqnarray}
F=\left(\frac{3\Omega_D
H^{2\delta-2}}{B}\right)^{{1}/{(4-2\delta)}}.
\end{eqnarray}
By substituting Eq.(\ref{dotPrho}) into the conservation law,
\begin{eqnarray}
\dot{\rho}_D+3H(1+\omega_D)\rho_D=0,
\end{eqnarray}
one finds the EoS parameter of THDE as
\begin{eqnarray}\label{PEoS1}
\omega_D=-1-\left(\frac{2\delta-4}{3}\right)(1+F).
\end{eqnarray}
Additionally, if we combine the time derivative of
$\Omega_{D}={\rho_{D}}/{(3m_{p}^{2}H^{2})}$ with Eqs. (\ref{7}),
(\ref{dotPrho}) and~(\ref{PEoS1}), then one may arrive at
\begin{eqnarray}\label{POmega1}
\Omega_{D}^{\prime}=\Omega_D(\Omega_D-1)[1+2F(\delta-2)-2\delta].
\end{eqnarray}
\begin{figure}[htp]
\begin{center}
\includegraphics[width=8cm]{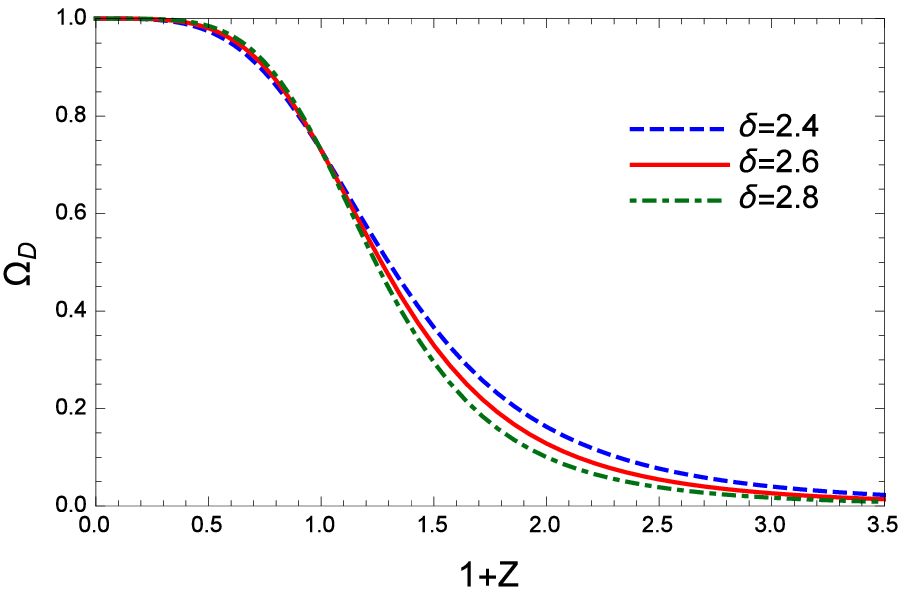}
\caption{The evolution of $\Omega_D$ versus redshift
parameter $z$ for non-interacting THDE with particle horizon as
the IR cutoff. Here, we have taken $\Omega^{0}_D=0.73$,
$B=2.4$ and $H(a=1)=67$. }\label{POmega-z4}
\end{center}
\end{figure}
\begin{figure}[htp]
\begin{center}
\includegraphics[width=8cm]{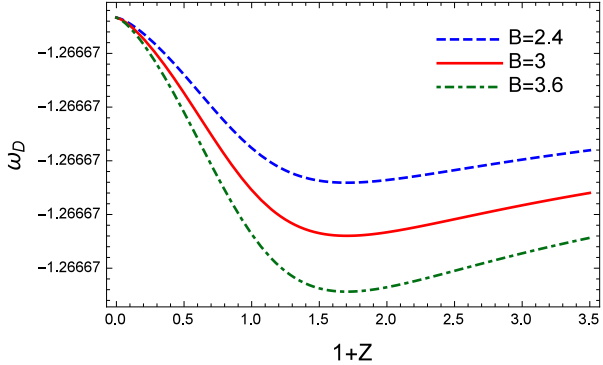}
\caption{The evolution of $\omega_D$ versus redshift
parameter $z$ for non-interacting THDE with particle horizon as
the IR cutoff . Here, we have taken $\Omega^{0}_D=0.73$,
$\delta=2.4$ and $H(a=1)=67$. }\label{Pw-z3}
\end{center}
\end{figure}
The deceleration parameter $q$ and the squared speed of sound
(defined in Eq.~(\ref{vs})), are also founded out as
\begin{eqnarray}\label{Pdeceleration1}
q=\frac{[1+(1-2F(\delta-2)-2\delta)\Omega_D]}{2}.
\end{eqnarray}
and
\begin{figure}[htp]
\begin{center}
\includegraphics[width=8cm]{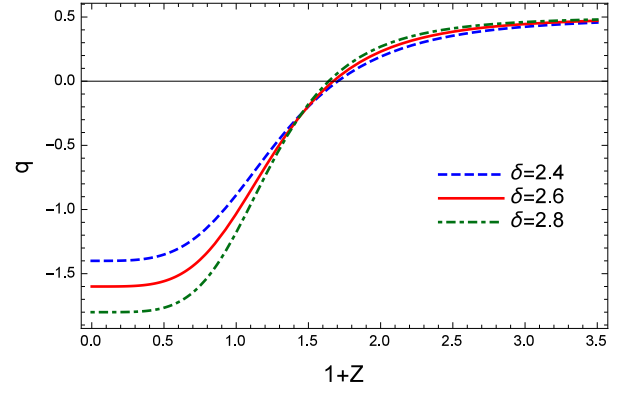}
\caption{The evolution of $q$ versus redshift parameter $z$
for non-interacting THDE with particle horizon as the IR
cutoff. Here, we have taken $\Omega^{0}_D=0.73$, $B=2.4$ and
$H(a=1)=67$. }\label{Pq-z4}
\end{center}
\end{figure}
\begin{eqnarray}
&&v_{s}^{2}=\frac{-2-9F-10F^{2}+4\delta(F+1)^{2}}{-6(F+1)}
\nonumber\\&&+\frac{F[-1+2F(-2+\delta)+2\delta]\Omega_D}{6(F+1)},
\end{eqnarray}
respectively. Here, in order to obtain Eq.~(\ref{Pdeceleration1}),
we employed Eq.~(\ref{PEoS1}) in writing Eq.~(\ref{7}), and then
we used relation~(\ref{q}). In
Figs.~\ref{POmega-z4}-\ref{Pvw3-z4}, the system parameters have
been plotted for some values of the system unknowns and the
initial condition $\Omega^{0}_D=0\cdot73$ and $H(a=1)=67$. As it
is apparent, although this cutoff leads to a model can provide
acceptable behavior for $\Omega_D$, $q$ and $\omega_D$, the model
is not stable.
%%%%%%%%%%%%%%%%%%%%%%%%%%%%%%%%%%%%%%%%%%%%%%%%%%%%%%%%%%%%%%
\subsection{Interacting}
\begin{figure}[htp]
\begin{center}
\includegraphics[width=8cm]{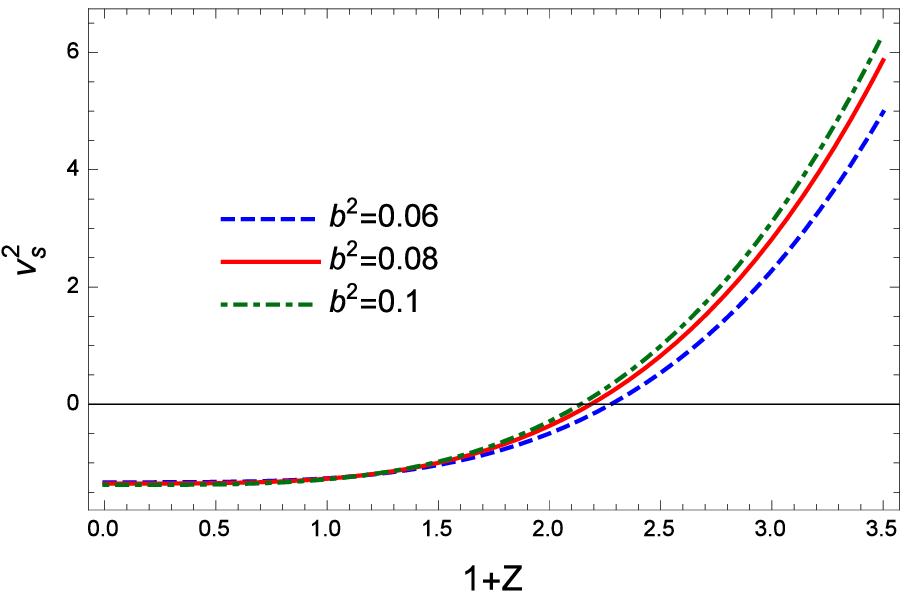}
\caption{The evolution of ${v}^{2}_{s}$ versus redshift
parameter $z$ for interacting THDE with particle horizon as
the IR cutoff. Here, we have taken $\Omega^{0}_D=0\cdot73$,
$B=2\cdot4$, $\delta=2\cdot4$ and $H(a=1)=67$.}\label{Pvs-z5}
\end{center}
\end{figure}

\begin{figure}[htp]
\begin{center}
\includegraphics[width=8cm]{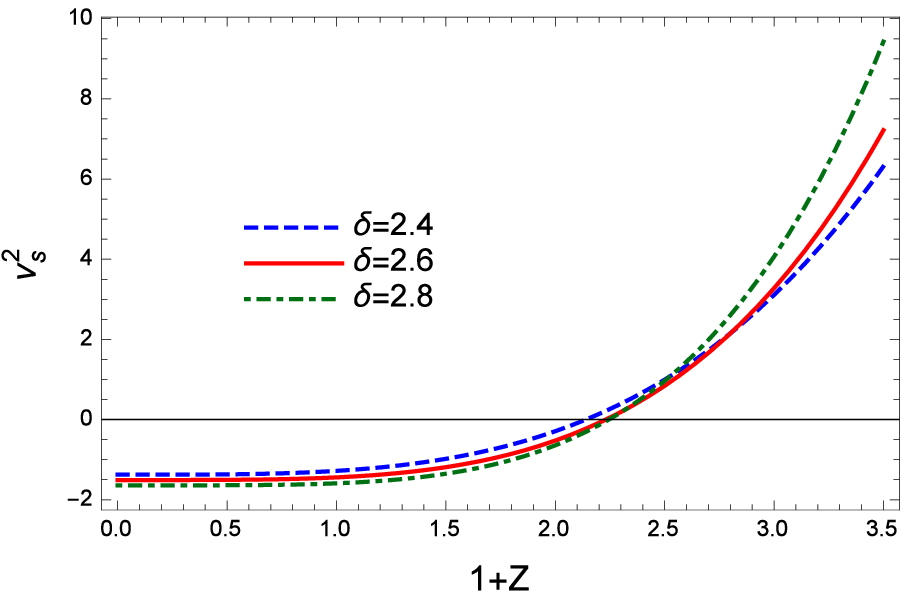}
\caption{The evolution of ${v}^{2}_{s}$ versus redshift
parameter $z$ for interacting THDE with particle horizon as
the IR cutoff. Here, we have taken $\Omega^{0}_D=0\cdot73$,
$b^2=0\cdot1$, $B=2\cdot4$ and $H(a=1)=67$.}\label{Pvs-z7}
\end{center}
\end{figure}

Using Eq.~(\ref{dotPrho}) and $Q=3b^2 H(\rho_m+\rho_D)$ in the
conservation equation (\ref{conD}), the EoS parameter is found as
\begin{eqnarray}\label{PEoS2}
\omega_D=-1-\frac{b^2}{\Omega_D}-\left(\frac{2\delta-4}{3}\right)(1+F).
\end{eqnarray}
For this choice of interaction, the evolution of density
parameter, the deceleration parameter $q$ and stability for the
model are calculated as
\begin{eqnarray}\label{POmega2}
\Omega_{D}^{\prime}=-3b^2
\Omega_D+\Omega_D(\Omega_D-1)[1+2F(\delta-2)-2\delta],
\end{eqnarray}
\begin{figure}[htp]
\begin{center}
\includegraphics[width=8cm]{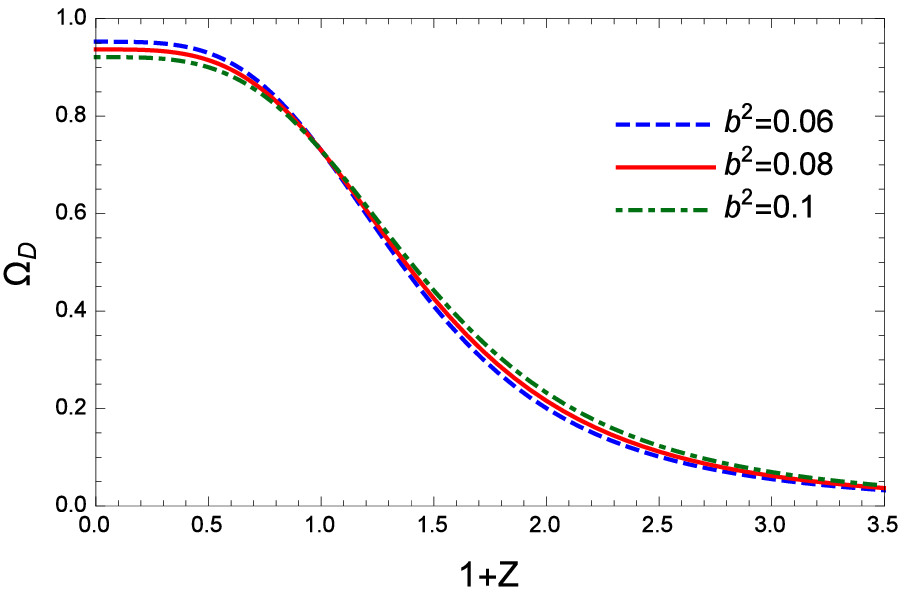}
\caption{The evolution of $\Omega_D$ versus redshift
parameter $z$ for interacting THDE with particle horizon as
the IR cutoff. Here, we have taken $\Omega^{0}_D=0\cdot73$,
$B=2\cdot4$, $\delta=2\cdot4$ and $H(a=1)=67$.}\label{POmega-z5}
\end{center}
\end{figure}
\begin{figure}[htp]
\begin{center}
\includegraphics[width=8cm]{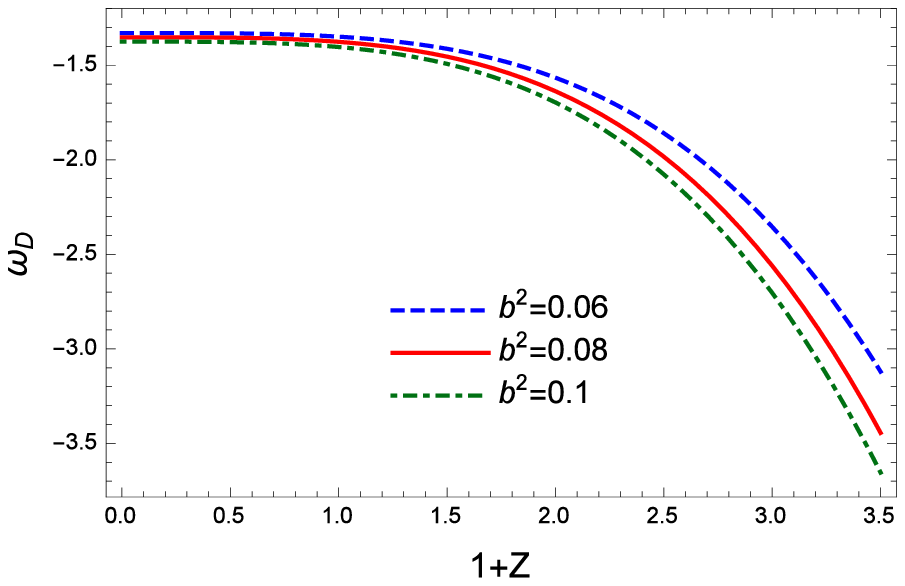}
\caption{The evolution of $\omega_D$ versus redshift
parameter $z$ for interacting THDE with particle horizon as
the IR cutoff. Here, we have taken $\Omega^{0}_D=0\cdot73$,
$B=2\cdot4$, $\delta=2\cdot4$ and $H(a=1)=67$.}\label{Pw-z5}
\end{center}
\end{figure}
\begin{eqnarray}\label{Pdeceleration2}
q=\frac{[1-3b^2+(1-2F(\delta-2)-2\delta)\Omega_D]}{2},
\end{eqnarray}
\begin{figure}[htp]
\begin{center}
\includegraphics[width=8cm]{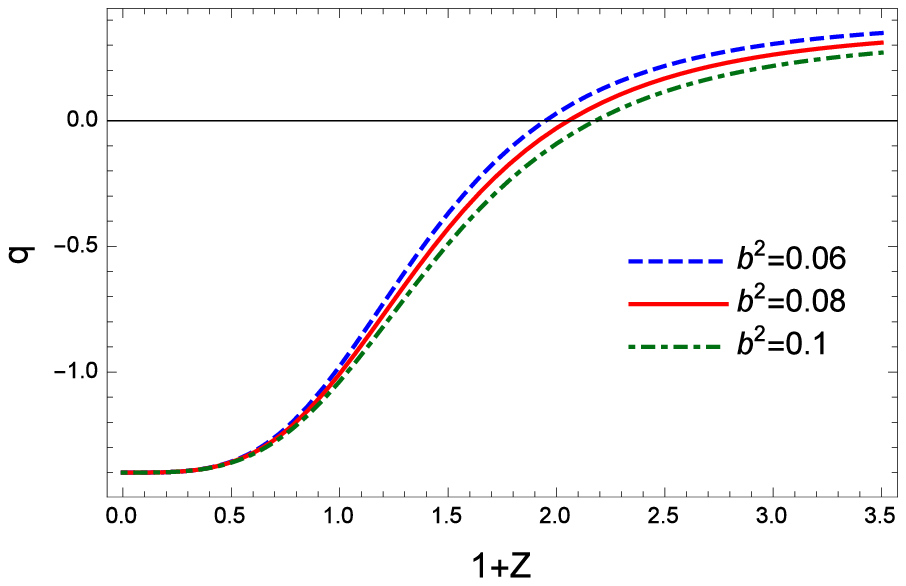}
\caption{The evolution of $q$ versus redshift parameter $z$
for interacting THDE with particle horizon as the IR
cutoff. Here, we have taken $\Omega^{0}_D=0\cdot73$, $B=2\cdot4$,
$\delta=2\cdot4$ and $H(a=1)=67$.}\label{Pq-z5}
\end{center}
\end{figure}
and
\begin{eqnarray}
&&v_{s}^{2}=\frac{9b^2(b^2-1)+\Omega_D[A-FB]}{-6(F+1)(\delta-2)\Omega_D},
\\&&A=(\delta-2)(-2-9F-10F^{2}+4(F+1)^{2}\delta)\nonumber\\&&+b^2(-3-6F+6\delta+3F\delta),\nonumber\\
&&B=[-1+2F(-2+\delta)+2\delta](\delta-2)\Omega_D,\nonumber
\end{eqnarray}
respectively. They are also plotted in
Figs.~\ref{POmega-z5}-\ref{Pvs-z7} for some values of the model's
parameters.
Fig.~\ref{Pvs-z7} and~\ref{Pvs-z5} show that, unlike the
noninteracting case, the model is stable for some values of $z$.
In addition, comparing Figs.~\ref{POmega-z5} and~\ref{POmega-z4}
with each other, we observe that the changes in the
density parameter of interacting case is slower than the
noninteracting case. Moreover, Figs.~\ref{Pw-z5} and~\ref{Pq-z5}
indicate that the model behaves as the phantom source, and thus,
the model eventually enters the accelerated phase with the
EoS for the universe being less than $-1$ (or equally $q<-1$).
%%%%%%%%%%%%%%%%%%%%%%%%%%%%%%%%%%%%%%%%%%%%%%%%%%%%%%%%%%%%%%%%%%%%%%%%%%%%%%
\section{THDE with GO horizon cutoff}\label{Interacting GO}
\subsection{Non-interacting}
In order to solve the causality and coincidence problems, Granda
and Oliveros (GO) \cite{GO1,GO2} suggested a new cutoff, usually
known as GO cutoff in the literatures, which is defined as
$L=(\gamma H^2+\zeta \dot{H})^{-{1}/{2}}$. In this case the energy
density of THDE  becomes
\begin{equation}\label{GOrho}
\rho_D=(\alpha H^2+\beta \dot{H})^{-\delta +2},
\end{equation}
which leads to
\begin{equation}\label{GOdotH}
\frac{\dot{H}}{H^2}= \frac{1}{\beta}\left(\frac{{(3m_p^2 \Omega_D)}^{\frac{1}
{2-\delta}}}{{H}^{\frac{2-2\delta}{2-\delta}}}-\alpha \right).
\end{equation}
Simple calculations for the deceleration and density parameters
yield
\begin{equation}\label{GOdeceleration}
q=-1-\frac{1}{\beta}\left(\frac{{(3m_p^2
\Omega_D)}^{\frac{1}{2-\delta}}}{{H}^{\frac{2-2\delta}{2-\delta}}}-\alpha
\right),
\end{equation}
and
\begin{equation}\label{GoOmega1}
\dot{\Omega}_D=\frac{\dot{\rho}_D}{3M_p^2H^2}-2\Omega_D\frac{\dot{H}}{H},
\end{equation}
respectively. For the $Q=0$ case, inserting the time derivative of
Eq.(\ref{frd}) into Eq.(\ref{conm}), we obtain
\begin{equation}\label{GoOmega2}
\frac{\dot{\rho}_D}{3m_p^2H^3}=\frac{2\dot{H}}{H^2}+3(1-\Omega_D).
\end{equation}
Combining with Eqs.~(\ref{GoOmega1}) and~(\ref{GOdotH}),
we obtain
\begin{equation}\label{GoOmega3}
{\Omega}^\prime_D=(1-\Omega_D)
\left[3+\frac{2}{\beta}\left(\frac{{(3m_p^2 \Omega_D)}^{\frac{1}{2-\delta}}}{{H}^{\frac{2-2\delta}{2-\delta}}}-\alpha \right)\right].
\end{equation}
In this manner, the EoS parameter of THDE is given by
\begin{equation}\label{GoEoS}
\omega_D=-1-\frac{1}{3\Omega_D}\left[\frac{2}{\beta}
\left(\frac{{(3m_p^2 \Omega_D)}^{\frac{1}{2-\delta}}}{{H}^{\frac{2-2\delta}{2-\delta}}}-\alpha \right)+3(1-\Omega_D)\right],
\end{equation}
where we have inserted $\dot{\rho}_D$ from
Eq.~(\ref{GoOmega2}) in the energy conservation law~(\ref{conD}).
Finally by taking time derivative from Eq.~(\ref{GoEoS}), and
using it in rewriting Eq.~(\ref{vs}), one can easily find
\begin{eqnarray}\nonumber
v_{s}^{2}=-1+\frac{2\alpha}{3\beta}+2\frac{3^{\frac{-1+\delta}{2-\delta}}~
{H}^{\frac{2-2\delta}{-2+\delta}} ~{\Omega_D}^{\frac{-1+\delta}{2-\delta}}}{\beta(-2+\delta)}~~~~~~~~~~~~~~~~~~~~~~~\\
-\frac{(2\alpha-3\beta)
~{H}^{\frac{2\delta}{-2+\delta}}(-1+\Omega_D)}{-2*{3}^{\frac{1}{2-\delta}}
~{H}^{\frac{2}{-2+\delta}}
~{\Omega_D}^{\frac{1}{2-\delta}}+ {H}^{\frac{2 \delta}{-2+\delta}}[2\alpha-3\beta+3\beta
\Omega_D]},
\end{eqnarray}
\begin{figure}[htp]
\begin{center}
\includegraphics[width=8cm]{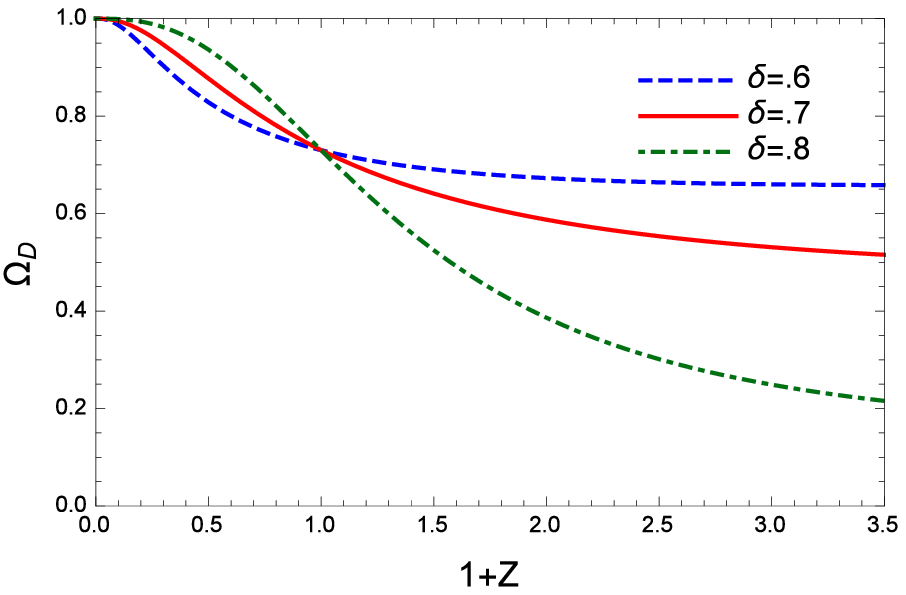}
\caption{The evolution of $\Omega_D$ versus redshift
parameter $z$ for noninteracting THDE with GO  horizon as
the IR cutoff. Here, we have taken
$\Omega^{0}_D=0\cdot73$, $\alpha=0\cdot8$, $\beta=0\cdot5$ and
$H(a=1)=67$.}\label{GoOmega-z1}
\end{center}
\end{figure}

\begin{figure}[htp]
\begin{center}
\includegraphics[width=8cm]{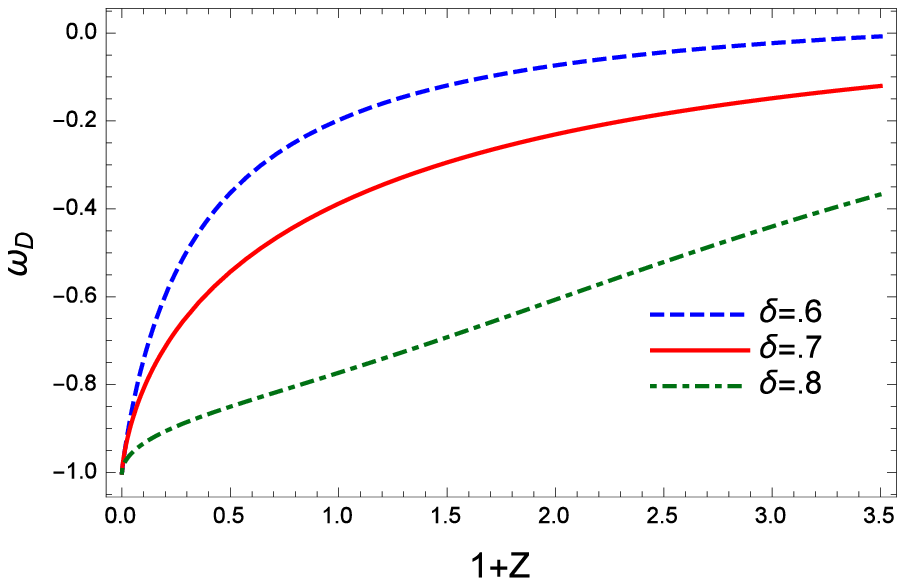}
\caption{The evolution of $\omega_D$ versus redshift
parameter $z$ for noninteracting THDE with GO horizon as
the IR cutoff. Here, we have taken
$\Omega^{0}_D=0\cdot73$, $\alpha=0\cdot8$, $\beta=0\cdot5$ and
$H(a=1)=67$.}\label{Gow-z1}
\end{center}
\end{figure}

\begin{figure}[htp]
\begin{center}
\includegraphics[width=8cm]{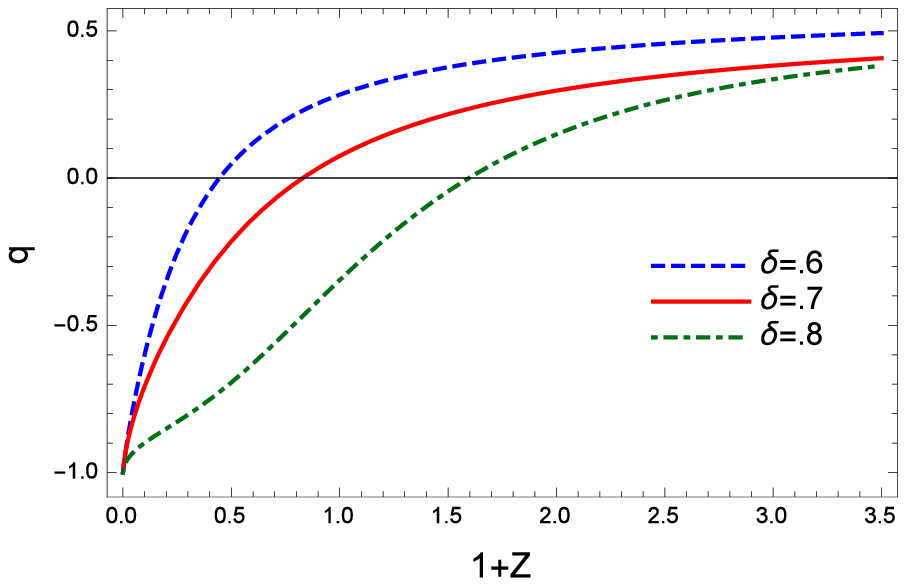}
\caption{The evolution of $q$ versus redshift parameter
$z$ for noninteracting THDE with GO horizon as the IR
cutoff. Here, we have taken
$\Omega^{0}_D=0\cdot73$,$\alpha=0\cdot8$, $\beta=0\cdot5$ and
$H(a=1)=67$.}\label{Goq-z1}
\end{center}
\end{figure}

\begin{figure}[htp]
\begin{center}
\includegraphics[width=8cm]{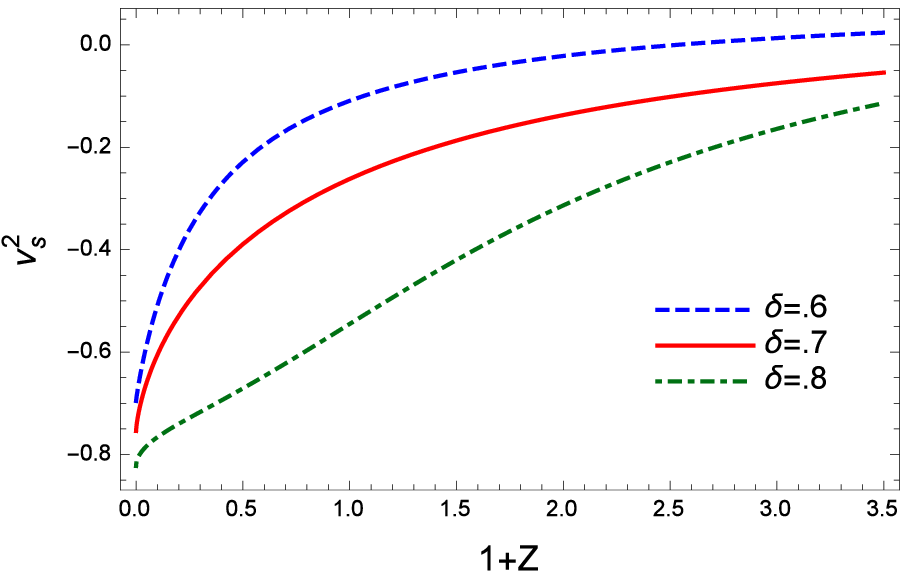}
\caption{The evolution of ${v}^{2}_{s}$ versus redshift
parameter $z$ for noninteracting HDE with GO horizon as
the IR cutoff. Here, we have taken
$\Omega^{0}_D=0\cdot73$, $\alpha=0\cdot8$, $\beta=0\cdot5$ and
$H(a=1)=67$.}\label{Govs-z1}
\end{center}
\end{figure}
In Figs.~\ref{GoOmega-z1}-\ref{Govs-z1}, the system parameters
including $\omega_D$, $q$, $\Omega_D$ and ${v}^{2}_{s}$
are plotted for some values of $\alpha$, $\beta$ and
$\delta$ by considering the initial conditions
$\Omega^{0}_D=0\cdot73$ and $H(a=1)=67$. It is interesting to note
here that the model begin to show stability from itself whenever
$q\rightarrow\frac{1}{2}$. Moreover, the depicted curves are some
of those which do not cross the phantom line for $z\geq-1$.
%%%%%%%%%%%%%%%%%%%%%%%%%%%%%%%%%%%%%%%%%%%%%%%%%%%%%%%%%%%%%%%%%%%%%%
\subsection{Interacting}
One can check that $q$ and $\omega_D$ have the same form as those
of the non-interacting case meaning that the mutual interaction
does not affect them. Thus, we only need the ${\Omega}^\prime_D$
and $v_{s}^{2}$ parameters evaluated as
\begin{equation}\label{GoOmega4}
{\Omega}^\prime_D=-3b^2+(1-\Omega_D)
\left[3+\frac{2}{\beta}\left(\frac{{(3m_p^2
\Omega_D)}^{\frac{1}{2-\delta}}}{{H}^{\frac{2-2\delta}{2-\delta}}}-\alpha
\right)\right],
\end{equation}
and
\begin{eqnarray}\nonumber
&&v_{s}^{2}=-1+\frac{2\alpha}{3\beta}+2\frac{3^{\frac{-1+\delta}{2-\delta}}~{H}^{\frac{2-2\delta}{-2+\delta}}
 ~{\Omega_D}^{\frac{-1+\delta}{2-\delta}}}{\beta(-2+\delta)}-~~~~~~~~~~~~~~~~~~~~~~~\\
&&\frac{(2\alpha-3\beta)
~{H}^{\frac{2\delta}{-2+\delta}}(-1+b^2+\Omega_D)}{-2{3}^{\frac{1}{2-\delta}}
~{H}^{\frac{2}{-2+\delta}} ~{\Omega_D}^{\frac{1}{2-\delta}}+
{H}^{\frac{2 \delta}{-2+\delta}}[2\alpha+3\beta(-1+b^2)+3\beta
\Omega_D]}, \nonumber\\
\end{eqnarray}
respectively.
\begin{figure}[htp]
\begin{center}
\includegraphics[width=8cm]{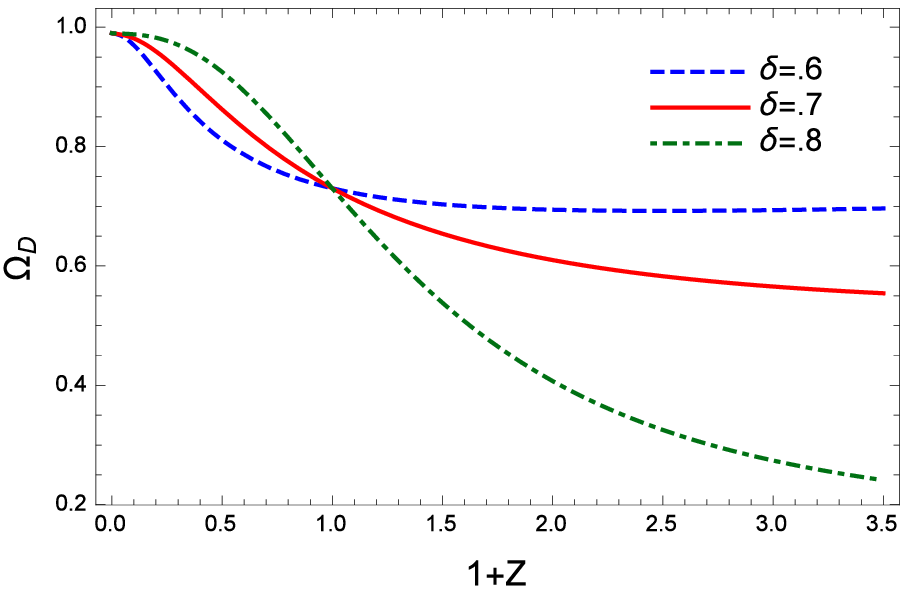}
\caption{The evolution of $\Omega_D$ versus redshift
parameter $z$ for
 interacting THDE with GO horizon as the
IR cutoff. Here, we have taken $\Omega^{0}_D=0\cdot73$,
$\alpha=0\cdot8$, $\beta=0\cdot5$, $b^2=0\cdot01$ and
$H(a=1)=67$.}\label{GoOmega-z2}
\end{center}
\end{figure}
\begin{figure}[htp]
\begin{center}
\includegraphics[width=8cm]{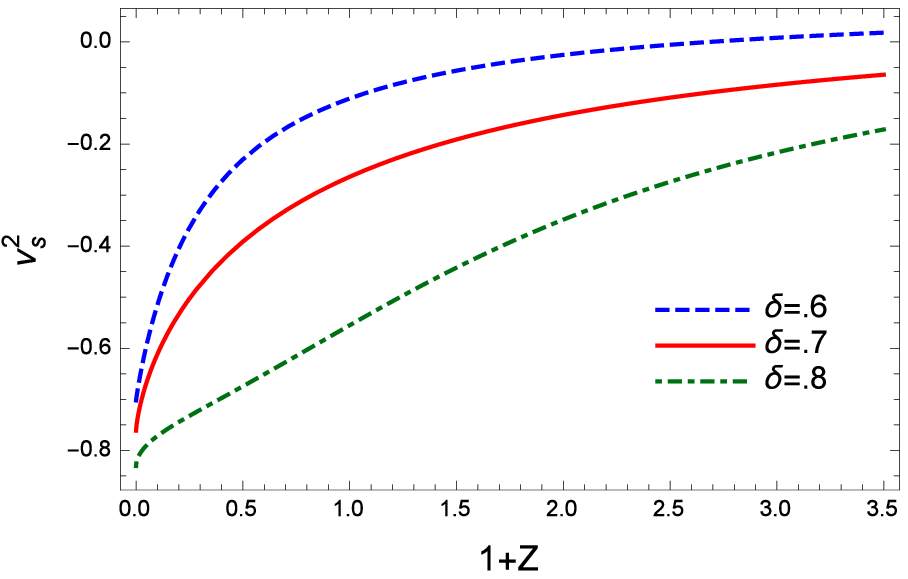}
\caption{The evolution of ${v}^{2}_{s}$ versus redshift
parameter $z$ for
 interacting THDE with GO horizon as the
IR cutoff. Here, we have taken $\Omega^{0}_D=0\cdot73$,
$\alpha=0\cdot8$, $\beta=0\cdot5$, $b^2=0\cdot01$ and
$H(a=1)=67$.}\label{Govs-z2}
\end{center}
\end{figure}
These parameters are plotted in
Figs.~\ref{GoOmega-z2}-\ref{Govs-z2} for some values of the system
constants. As it is apparent, for $q\rightarrow {1}/{2}$ the model
shows stability from itself, a result in full agreement with the
noninteracting case.
%%%%%%%%%%%%%%%%%%%%%%%%%%%%%%%%%%%%%%%%%%%%%%%%%%%%%%%%%%%%%%%%%%%%%%%%%%%%%%%%%%%%%%%%%%%%%%%%
\section{HDE with Ricci horizon cutoff}\label{Interacting Ricci}
\subsection{Non-interacting}
The energy density of THDE with Ricci scalar as the IR cutoff is
written as \cite{Gao}
\begin{equation}\label{Rrho}
\rho_D=\lambda(2H^2+ \dot{H})^{-\delta +2},
\end{equation}
where $\lambda$ is an unknown HDE constant as usual
\cite{Li2004,Gao}. This energy density is also obtainable by
inserting $\alpha=2\beta$ in Eq.~(\ref{GOrho}) and defining new
unknown constant $\lambda$ as $\lambda=\beta^{2-\delta}$, a
desired result. In order to find the deceleration parameter $q$,
we rewrite Eq.(\ref{Rrho}) as
\begin{equation}\label{RdotH}
\frac{\dot{H}}{H^2}= \left(\frac{{(3 {\lambda}^{-1} m_p^2 \Omega_D)}^{\frac{1}
{2-\delta}}}{{H}^{\frac{2-2\delta}{2-\delta}}}-2 \right),
\end{equation}
and use Eq.~(\ref{q}) to obtain
\begin{equation}\label{Rdeceleration}
q=-1-\left(\frac{{(3 {\lambda}^{-1} m_p^2 \Omega_D)}^{\frac{1}{2-\delta}}}
{{H}^{\frac{2-2\delta}{2-\delta}}}-2 \right).
\end{equation}
It is also a matter of calculation to combine
Eqs.~(\ref{GoOmega1}) and~(\ref{GoOmega2}) with~(\ref{RdotH}) to
reach at
\begin{equation}\label{ROmega}
{\Omega}^\prime_D=(1-\Omega_D)
\left[3+2\left(\frac{{(3 {\lambda}^{-1} m_p^2 \Omega_D)}^{\frac{1}{2-\delta}}}
{{H}^{\frac{2-2\delta}{2-\delta}}}-2 \right)\right].
\end{equation}
In this manner, we have
\begin{equation}\label{REoS}
\omega_D=-1-\frac{1-\Omega_D}{\Omega_D}[\frac{2}{3(1-\Omega_D)}(\frac{{(3
{\lambda}^{-1} m_p^2
\Omega_D)}^{\frac{1}{2-\delta}}}{{H}^{\frac{2-2\delta}{2-\delta}}}-2
)+1],
\end{equation}
where Eqs.~(\ref{GoOmega2}),~(\ref{conD}) and~(\ref{RdotH}) have
been used to obtain the above result. Moreover, by taking
time derivative from Eq.~(\ref{REoS}) we find
\begin{eqnarray}
&&v_{s}^{2}=\frac{1}{3}+2\frac{{3}^{\frac{-1+\delta}{2-\delta}}~
{H}^{\frac{2-2\delta}{-2+\delta}} ~{( {\lambda}^{-1}  \Omega_D)}^{\frac{1}{2-\delta}}}{\Omega_D(-2+\delta)}\\
&&+\frac{(-1+\Omega_D)}{-1-3\Omega_D+2\times{3}^{\frac{1}{2-\delta}}
~{H}^{\frac{2-2\delta}{-2+\delta}}
~{( {\lambda}^{-1}\Omega_D)}^{\frac{1}{2-\delta}}}.\nonumber
\end{eqnarray}
\begin{figure}[htp]
\begin{center}
\includegraphics[width=8cm]{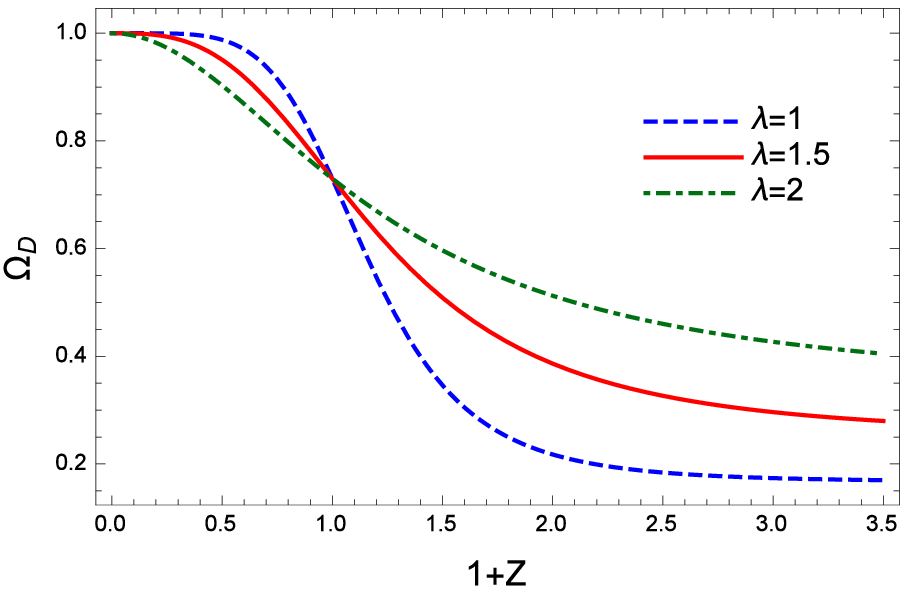}
\caption{The evolution of $\Omega_D$ versus redshift
parameter $z$ for non-interacting THDE with Ricci horizon as
the IR cutoff. Here, we have taken $\Omega^{0}_D=0\cdot73$
and $H(a=1)=67$ and $\delta=1$.}\label{ROmega-z1}
\end{center}
\end{figure}
\begin{figure}[htp]
\begin{center}
\includegraphics[width=8cm]{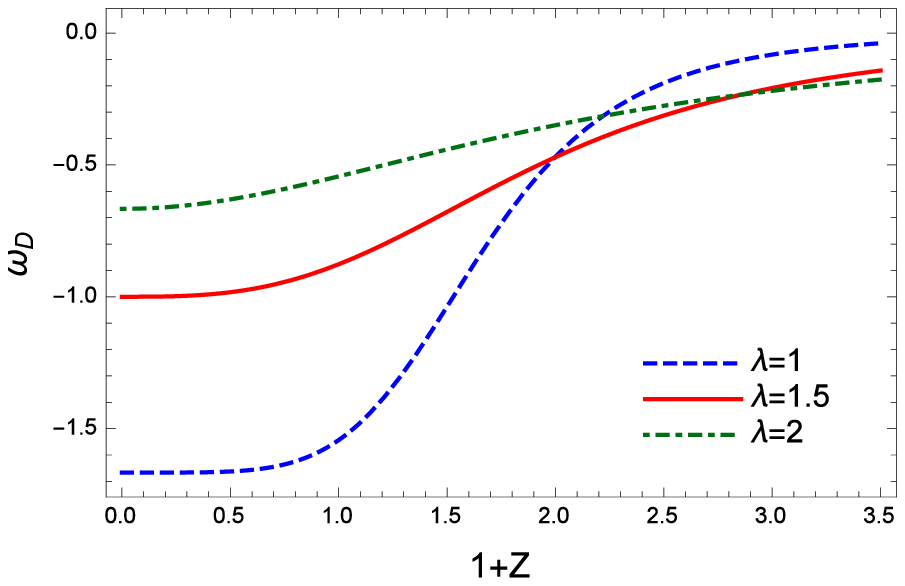}
\caption{The evolution of $\omega_D$ versus redshift
parameter $z$ for non-interacting THDE with Ricci horizon as
the IR cutoff. Here, we have taken $\Omega^{0}_D=0\cdot73$
and $H(a=1)=67$ and $\delta=1$.}\label{Rw-z1}
\end{center}
\end{figure}
\begin{figure}[htp]
\begin{center}
\includegraphics[width=8cm]{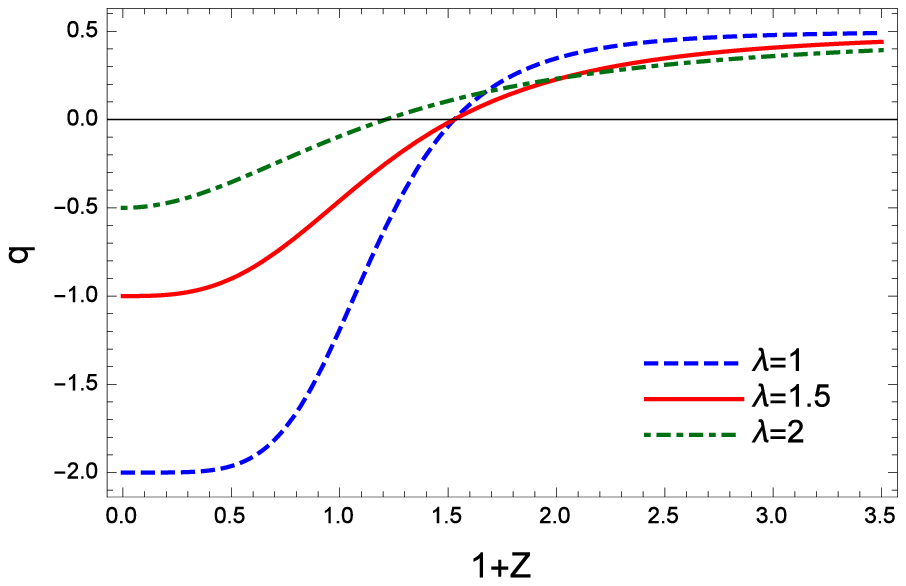}
\caption{The evolution of $q$ versus redshift parameter $z$
for non-interacting THDE with Ricci horizon as the IR
cutoff. Here, we have taken $\Omega^{0}_D=0\cdot73$ and $H(a=1)=67$
and $\delta=1$.}\label{Rq-z1}
\end{center}
\end{figure}
\begin{figure}[htp]
\begin{center}
\includegraphics[width=8cm]{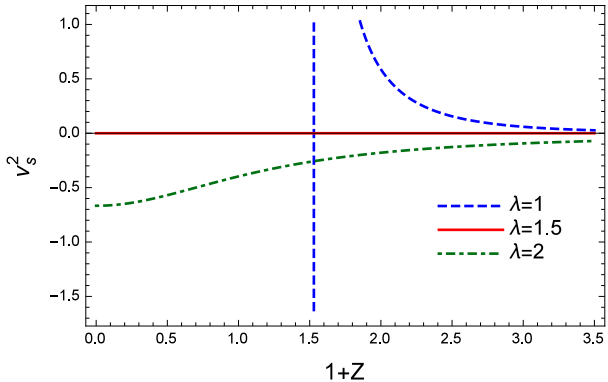}
\caption{The evolution of ${v}^{2}_{s}$ versus redshift
parameter $z$ for non-interacting THDE with Ricci horizon as
the IR cutoff. Here, we have taken $\Omega^{0}_D=0\cdot73$
and $H(a=1)=67$ and $\delta=1$.}\label{Rvs-z1}
\end{center}
\end{figure}
It can be seen from Figs.~\ref{ROmega-z1}-\ref{Rvs-z1} that the
current accelerated universe can be achieved. The $\lambda=1$ case
is interesting, because unlike SMHDE \cite{Sayahian}, this model
is stable (unstable) for $q>0$ ($q<0$). Hence, since the Ricci
horizon is a special case of the GO cutoff, the GO cutoff can also
produce the same results if proper values for the system unknown
constants have been chosen.
%%%%%%%%%%%%%%%%%%%%%%%%%%%%%%%%%%%%%%%%%%%%%%%%%%%%%%%%%%%%%%%%%%%55
\subsection{Interacting}
\begin{figure}[htp]
\begin{center}
\includegraphics[width=8cm]{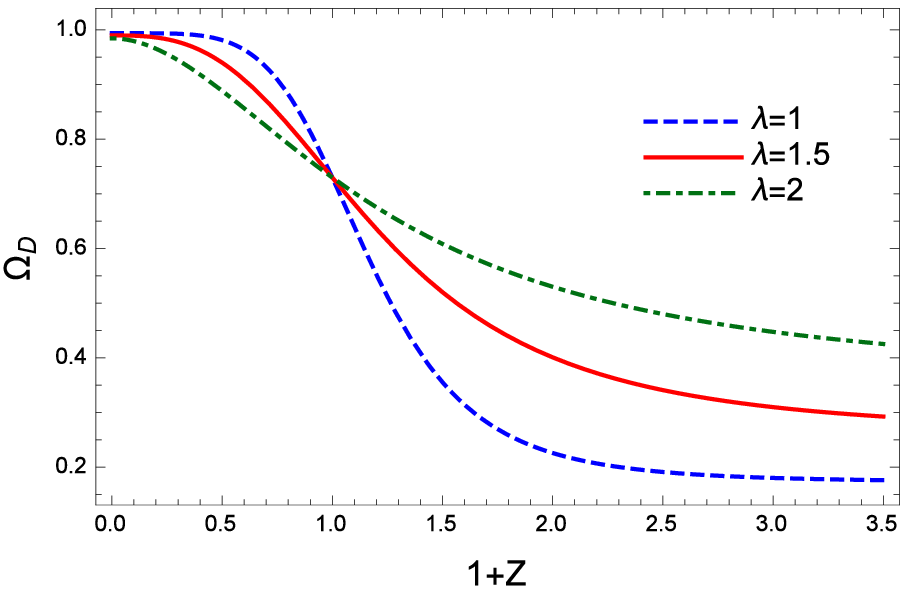}
\caption{The evolution of $\Omega_D$ versus redshift
parameter $z$ for interacting THDE with Ricci horizon as
the IR cutoff. Here, we have taken $\Omega^{0}_D=0\cdot73$,
$H(a=1)=67$, $b^2=0\cdot01$ and $\delta=1$.}\label{ROmega-z2}
\end{center}
\end{figure}

\begin{figure}[htp]
\begin{center}
\includegraphics[width=8cm]{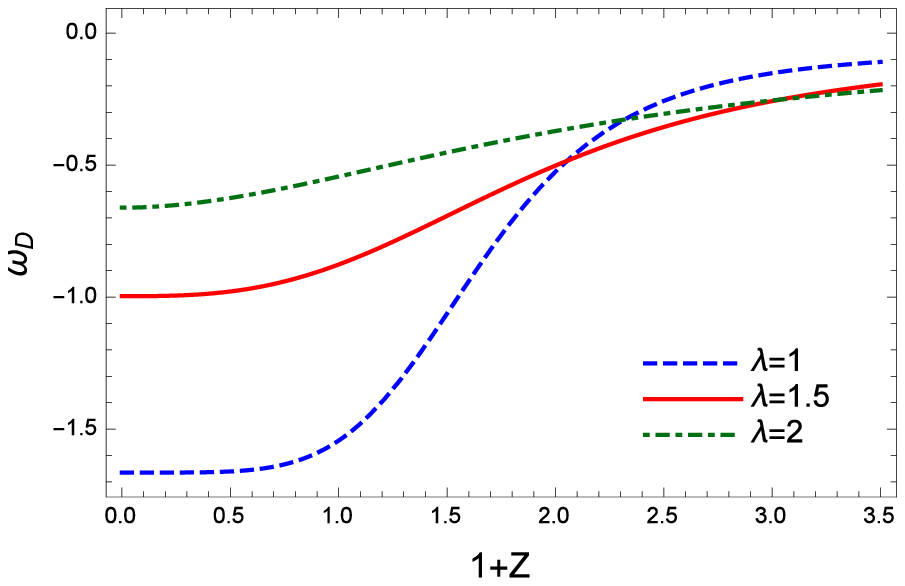}
\caption{The evolution of $\omega_D$ versus redshift
parameter $z$ for interacting THDE with Ricci horizon as
the IR cutoff. Here, we have taken $\Omega^{0}_D=0\cdot73$
and $H(a=1)=67$, $b^2=0\cdot01$ and $\delta=1$.}\label{Rw-z2}
\end{center}
\end{figure}

\begin{figure}[htp]
\begin{center}
\includegraphics[width=8cm]{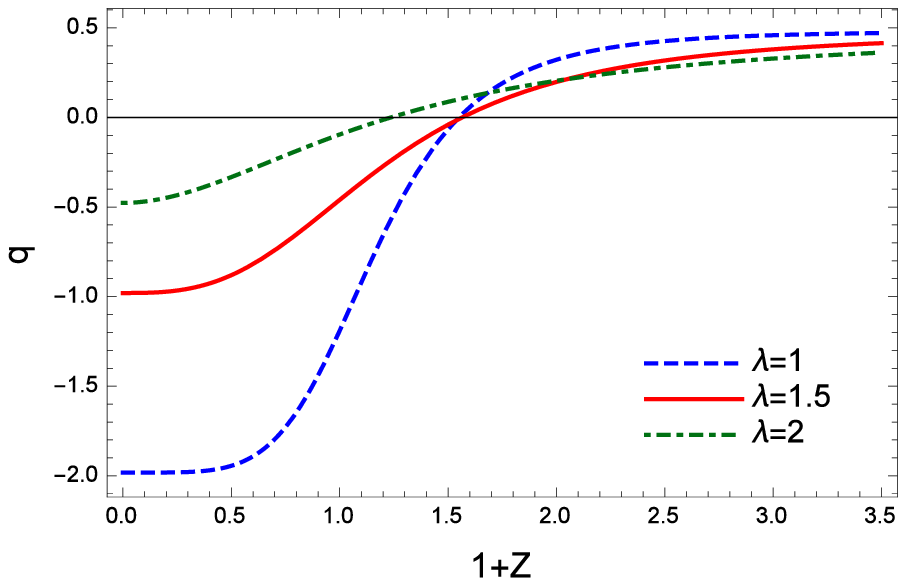}
\caption{The evolution of $q$ versus redshift parameter $z$
for interacting THDE with Ricci horizon as the IR cutoff.
Here, we have taken $\Omega^{0}_D=0\cdot73$ and $H(a=1)=67$,
$b^2=0\cdot01$ and $\delta=1$.}\label{Rq-z2}
\end{center}
\end{figure}

\begin{figure}[htp]
\begin{center}
\includegraphics[width=8cm]{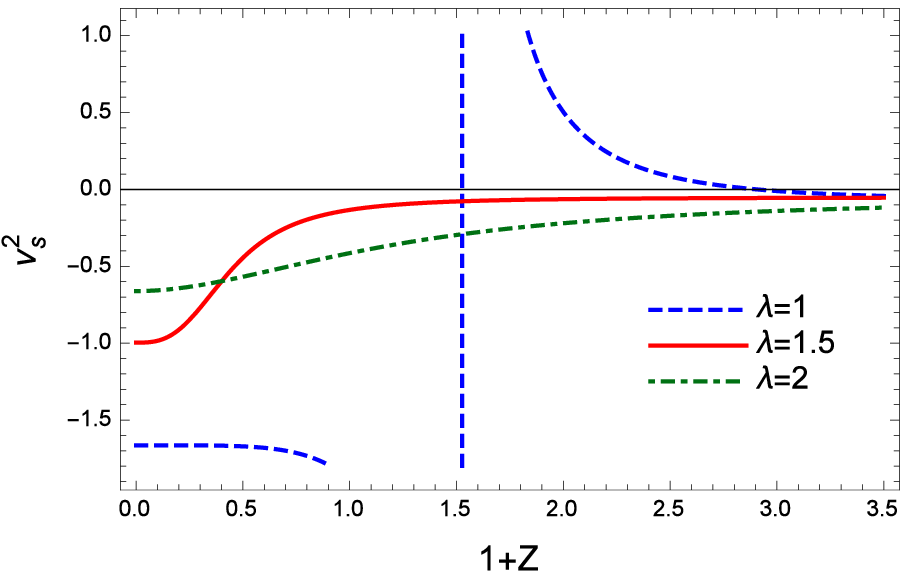}
\caption{The evolution of ${v}^{2}_{s}$ versus redshift
parameter $z$ for interacting THDE with Ricci horizon as
the IR cutoff. Here, we have taken $\Omega^{0}_D=0\cdot73$,
$H(a=1)=67$, $b^2=0\cdot01$ and $\delta=1$.}\label{Rvs-z2}
\end{center}
\end{figure}

Just the same as the GO cutoff, one can easily check that we only
need to calculate the $\Omega_D$ and $v_{s}^{2}$ parameters in
this case, a result due to the fact that the Ricci cutoff is a
special case of the GO cutoff. The calculations lead to
\begin{equation}\label{ROmega1}
{\Omega}^\prime_D=-3b^2+(1-\Omega_D)[3+2(\frac{{(3 {\lambda}^{-1}
m_p^2
\Omega_D)}^{\frac{1}{2-\delta}}}{{H}^{\frac{2-2\delta}{2-\delta}}}-2)],
\end{equation}
and
\begin{eqnarray}
&&v_{s}^{2}=\frac{1}{3}+\frac{2*{3}^{\frac{-1+\delta}{2-\delta}}~{H}^{\frac{2-2\delta}{-2+\delta}}
~{( {\lambda}^{-1}  \Omega_D)}^{\frac{1}{2-\delta}}}{\Omega_D(-2+\delta)}\\
&&+\frac{(-1+b^2+\Omega_D)}{-1-3b^2-3\Omega_D+2*{3}^{\frac{1}{2-\delta}}
~{H}^{\frac{2-2\delta}{-2+\delta}}
~{( {\lambda}^{-1}\Omega_D)}^{\frac{1}{2-\delta}}},\nonumber
\end{eqnarray}
In Figs.\ref{ROmega-z2}-\ref{Rvs-z2}, the system parameters have
been plotted versus $z$ for some values of the unknown constants.
It is obvious that the system parameters affected by the mutual
interaction. It is also interesting to note that while $v_{s}^{2}$
was not negative for $\lambda=1\cdot5$ in the non-interacting
case, here, we always have it is not stable for all values of
$v_{s}^{2}<0$.
%%%%%%%%%%%%%%%%%%%%%%%%%%%%%%%%%%%%%%%%%%%%%%%%%%%%%%%%%%%%%%%%%%%%%%%%%%%%%%
\section{Closing remarks}
In the shadow of the holographic principle and based on the
non-additive generalized Tsallis entropy expression
\cite{Tsallis}, a new holographic dark energy model called THDE
has recently been proposed \cite{THDE}. In this
paper, by considering various IR cutoffs, including the particle
horizon, the Ricci horizon and the GO cutoff in
the background of the FRW universe, we investigated the
evolution of the THDE models and studied their
cosmological consequences. We found out that when the particle
horizon is considered as IR cutoff, then the THDE model can
explain the current acceleration of the universe expansion. This
is in contrast to the usual HDE model which cannot lead to an
accelerated universe, if one consider the particle horizon as IR
cutoff \cite{Li2004}. We also explored the sound stability of the
THDE models with various cutoffs. In this manner, the assumed
mutual interaction between the cosmos sectors makes the model to
be stable for some values of the redshift parameter $z$. For the
GO and the Ricci horizon cutoffs, we found out that
although acceptable behavior for some parameters of the
system, including $q$, the density parameter and
$\omega_D$, are achievable, the model is not always stable.
Finally, we have explored the effects of considering a
mutual interaction between the two dark sectors of the universe on
the behavior of the solutions.
%%%%%%%%%%%%%%%%%%%%%%%%%%%%%%%%%%%%%%%%%%%%%%%%%%%%%%%%%%%%%%%%%%%%%%%%%%%%%%%%%%
\acknowledgments{We thank Shiraz University Research Council. This
work has been supported financially by Research Institute for
Astronomy \& Astrophysics of Maragha (RIAAM), Iran. The work of KB
was supported in part by the JSPS KAKENHI Grant Number JP 25800136
and Competitive Research Funds for Fukushima University Faculty
(18RI009).}
%%%%%%%%%%%%%%%%%%%%%%%%%%%%%%%%%%%%%%%%%%%%%%%%%%%%%%%%%%%%%%%%%%%%%%%%%%%%%%%%%%%%%


\begin{thebibliography}{99}

\bibitem{THDE} M. Tavayef, A. Sheykhi, K. Bamba and H. Moradpour, Phys. Lett. B. {\bf781}, 195 (2018).
\bibitem{Riess} A. G. Riess\textit{ et al}., Astron. J. {\bf 116}, 1009 (1998).
\bibitem{Riess1} S. Perlmutter\textit{ et al}., Astrophys. J.  {\bf 517},  565 (1999).
\bibitem{Riess2} P. deBernardis, \textit{et al}., Nature {\bf 404}, 955 (2000).
\bibitem{Riess3} S. Perlmutter,\textit{et al}., Astrophys. J. {\bf 598}, 102 (2003).
\bibitem{COL2001} M. Colless \textit{et al}., Mon. Not. R. Astron. Soc. \textbf{328}, 1039 (2001).
\bibitem{COL20011} M. Tegmark \textit{etal}., Phys. Rev. D \textbf{69}, 103501 (2004).
\bibitem{COL20012} S. Cole\textit{ et al}., Mon. Not. R. Astron. Soc. \textbf{362}, 505 (2005).
\bibitem{COL20013} V. Springel, C. S. Frenk, and S. M. D. White, Nature(London) \textbf{440}, 1137 (2006).
\bibitem{Ade2014}P.A.R. Ade et al., Astron. Astrophys. {\bf571}, A16 (2014).
\bibitem{Astier} Astier, P., et al., Astron. Astrophys. {\bf447}, 31 (2006).
\bibitem{Astier2006} Riess, A.G., et al., Astrophys. J. {\bf659}, 98 (2007).
\bibitem{Astier20061} Spergel, D.N., et al., Astrophys. J. Suppl. Ser. {\bf148}, 175 (2003).
\bibitem{Astier20062} Peiris, H.V., et al., Astrophys. J. Suppl. Ser. {\bf148}, 213 (2003).
\bibitem{Astier20063} Spergel, D.N., et al., Astrophys. J. Suppl. Ser. {\bf170}, 377 (2007).
\bibitem{Astier20064} Komatsu, E., et al., [arXiv:0803.0547].
\bibitem{R-DE-MG}
%
%\cite{Nojiri:2010wj}
%\bibitem{Nojiri:2010wj}
S.~Nojiri and S.~D.~Odintsov,
%``Unified cosmic history in modified gravity: from F(R)
%theory to Lorentz non-invariant models,''
Phys.\ Rept.\ {\bf 505}, 59 (2011);\\
%[arXiv:1011.0544 [gr-qc]];\\
%%CITATION = ARXIV:1011.0544;%%
%
%\cite{Nojiri:2006ri}
%\bibitem{Nojiri:2006ri}
S.~Nojiri and S.~D.~Odintsov,
%``Introduction to modified gravity and gravitational alternative
%for dark energy,''
eConf C {\bf 0602061} (2006) 06
[Int.\ J.\ Geom.\ Meth.\ Mod.\ Phys.\ {\bf 4}, 115 (2007)];\\
%[hep-th/0601213];\\
%%CITATION = HEP-TH/0601213;%%
%
%\bibitem{Book-Capozziello-Faraoni}
S.~Capozziello and V.~Faraoni,
\textit{Beyond Einstein Gravity}
(Springer, Dordrecht, 2010);\\
%
%\cite{Capozziello:2011et}
%\bibitem{Capozziello:2011et}
S.~Capozziello and M.~De Laurentis,
%``Extended Theories of Gravity,''
Phys.\ Rept.\ {\bf 509}, 167 (2011);\\
%[arXiv:1108.6266 [gr-qc]];\\
%%CITATION = ARXIV:1108.6266;%%
%
%\cite{Bamba:2012cp}
%\bibitem{Bamba:2012cp}
  K.~Bamba, S.~Capozziello, S.~Nojiri and S.~D.~Odintsov,
  %``Dark energy cosmology: the equivalent description via
  %different theoretical models and cosmography tests,''
  Astrophys.\ Space Sci.\  {\bf 342}, 155 (2012);\\
%  [arXiv:1205.3421 [gr-qc]]; \\
  %%CITATION = ARXIV:1205.3421;%%
  %324 citations counted in INSPIRE as of 14 Jan 2015
%
%\cite{Joyce:2014kja}
%\bibitem{Joyce:2014kja}
  A.~Joyce, B.~Jain, J.~Khoury and M.~Trodden,
  %``Beyond the Cosmological Standard Model,''
  Phys.\ Rept.\  {\bf 568}, 1 (2015);\\
%  [arXiv:1407.0059 [astro-ph.CO]];\\
  %%CITATION = ARXIV:1407.0059;%%
  %70 citations counted in INSPIRE as of 15 Apr 2015
%
%\cite{Koyama:2015vza}
%\bibitem{Koyama:2015vza}
  K.~Koyama,
  %``Cosmological Tests of Modified Gravity,''
  Rept.\ Prog.\ Phys.\  {\bf 79}, 046902 (2016);\\
%  doi:10.1088/0034-4885/79/4/046902
%  [arXiv:1504.04623 [astro-ph.CO]];\\
  %%CITATION = doi:10.1088/0034-4885/79/4/046902;%%
  %62 citations counted in INSPIRE as of 27 Jun 2016
%
%\cite{Bamba:2015uma}
%\bibitem{Bamba:2015uma}
  K.~Bamba and S.~D.~Odintsov,
  %``Inflationary cosmology in modified gravity theories,''
  Symmetry {\bf 7}, 220 (2015);\\
%  doi:10.3390/sym7010220
%  [arXiv:1503.00442 [hep-th]].
  %%CITATION = doi:10.3390/sym7010220;%%
  %39 citations counted in INSPIRE as of 04 Mar 2016
%
%\cite{Cai:2015emx}
%\bibitem{Cai:2015emx}
  Y.~F.~Cai, S.~Capozziello, M.~De Laurentis and E.~N.~Saridakis,
  %``f(T) teleparallel gravity and cosmology,''
  Rept.\ Prog.\ Phys.\  {\bf 79}, 106901 (2016)
%  doi:10.1088/0034-4885/79/10/106901
%  [arXiv:1511.07586 [gr-qc]].
  %%CITATION = doi:10.1088/0034-4885/79/10/106901;%%
  %212 citations counted in INSPIRE as of 12 Jun 2018
%
%\cite{Nojiri:2017ncd}
%\bibitem{Nojiri:2017ncd}
  S.~Nojiri, S.~D.~Odintsov and V.~K.~Oikonomou,
  %``Modified Gravity Theories on a Nutshell: Inflation,
  %Bounce and Late-time Evolution,''
  Phys.\ Rept.\  {\bf 692}, 1 (2017).
%  doi:10.1016/j.physrep.2017.06.001
%  [arXiv:1705.11098 [gr-qc]].
  %%CITATION = doi:10.1016/j.physrep.2017.06.001;%%
  %132 citations counted in INSPIRE as of 16 Mar 2018
%
\bibitem{Hooft} G. t Hooft, gr-qc/9310026.
\bibitem{Hooft1995} L. Susskind, J. Math. Phys. {\bf36}, 6377 (1995).
\bibitem{Cohen} A. Cohen, D. Kaplan, A. Nelson, Phys. Rev. Lett. 82 (1999)
4971.
\bibitem{Zhang2005} X. Zhang, F. Q. Wu, Phys. Rev. D {\bf72}, 043524 (2005).
\bibitem{Zhang20051} X. Zhang, F. Q. Wu, Phys. Rev. D {\bf76}, 023502 (2007).
\bibitem{Zhang20052} Q. G. Huang, Y.G. Gong, JCAP {\bf0408}, 006 (2004).
\bibitem{Zhang20053} K. Enqvist, S. Hannestad, M. S. Sloth, JCAP {\bf0502}, 004 (2005).
\bibitem{Zhang20054} J. Y. Shen, B. Wang, E. Abdalla, R.K. Su, Phys. Lett. B {\bf609}, 200 (2005).
\bibitem{Li2004} M. Li, Phys. Lett. B {\bf603}, 1 (2004);\\
Q.G. Huang, M. Li, JCAP {\bf0408} 013 (2004).
\bibitem{Hsu} S. D. H. Hsu, Phys. Lett. B {\bf 594}, 13 (2004).


\bibitem{Nojiri:2005pu}
  S.~Nojiri and S.~D.~Odintsov,
  %``Unifying phantom inflation with late-time acceleration:
  %Scalar phantom-non-phantom transition model and generalized holographic
  %dark energy,''
  Gen.\ Rel.\ Grav.\  {\bf 38}, 1285 (2006);\\
%  doi:10.1007/s10714-006-0301-6
%  [hep-th/0506212].
%
%\cite{Nojiri:2017opc}
%\bibitem{Nojiri:2017opc}
  S.~Nojiri and S.~D.~Odintsov,
  %``Covariant Generalized Holographic Dark Energy and Accelerating Universe,''
  Eur.\ Phys.\ J.\ C {\bf 77}, 528 (2017).
%  doi:10.1140/epjc/s10052-017-5097-x
%  [arXiv:1703.06372 [hep-th]].
%
\bibitem{GO1} L. N. Granda, A. Oliveros, Phys. Lett. B {\bf669}, 275 (2008).
\bibitem{GO2} L. N. Granda, A. Oliveros, Phys. Lett. B {\bf671275}, 199 (2009).
\bibitem{SSB} M. Sharif, Syed. Asif Ali Shah and K. Bamba,
  %``New Holographic Dark Energy Model in Brans-Dicke Theory,''
  Symmetry {\bf 2018}, 153 (2018).
%  doi.org/10.3390/sym10050153


\bibitem{Wang1} B. Wang, Y. Gong and E. Abdalla, Phys. Lett. B \textbf{624}, 141 (2005);\\
B. Wang, C. Y. Lin and E. Abdalla, Phys. Lett. B \textbf{637}, 357 (2005);\\
M. R. Setare, Phys. Lett. B \textbf{642},1 (2006).

\bibitem{Wang2} B. Wang, E. Abdalla, R. K. Su, Phys. Lett. B 611 (2005)
21;\\
J. Y. Shen, B. Wang, E. Abdalla, R. K. Su, Phys. Lett.
B {\bf609}, 200 (2005);\\
C. Feng, B. Wang, Y. Gong, R. K. Su, JCAP {\bf 0709}, 005 (2007)  ;\\
B. Wang, C. Y. Lin. D. Pavon and E. Abdalla, Phys. Lett. B {\bf
662}, 1 (2008).

\bibitem{cop} E. J. Copeland, M. Sami and S. Tsujikawa, Int. J. Mod. Phys. D
{\bf15},  1753 (2006).

\bibitem{Zim} W. Zimdahl and D. Pavon, Classical Quantum Gravity {\bf24}, 5461
(2007).

\bibitem{shey0} A. Sheykhi, Phys. Lett. B {\bf681}, 205 (2009);\\  A. Sheykhi, Phys. Rev.  D {\bf84}, 107302
(2011).
\bibitem{shey1} A. Sheykhi, Class. Quantum Grav. {\bf27}, 025007
(2010);\\ A. Sheykhi, Mubasher Jamil, Phys. Lett. B {\bf694}, 284
(2011);\\M. Jamil, K. Karami, A. Sheykhi, E. Kazemi, and Z.
Azarmi, Int. J. Theor. Phys. \textbf{51}, 604 (2012);\\ A. Sheykhi
et al., Gen. Relativ. Gravit. \textbf{44}, 623 (2012); A. Sheykhi,
M.S. Movahed, E. Ebrahimi, Astrophys. Space Sci. {\bf339} 93
(2012).

\bibitem{shey2} A. Sheykhi, M. Jamil, Gen Relativ Gravit {\bf43}, 2661
(2011);\\S. Ghaffari, M. H. Dehghani, and A. Sheykhi, Phys. Rev. D
\textbf{89}, 123009 (2014).


\bibitem{Majhi} A. Majhi, Phys. Lett. B {\bf775}, 32 (2017).

\bibitem{Abe} S. Abe, Phys. Rev. E {\bf63}, 061105 (2001);\\
 H. Touchette, Physica A {\bf305}, 84 (2002).


\bibitem{Sayahian} A. Sayahian Jahromi et al, Phys. Lett. B {\bf780}, 21 (2018).
\bibitem{Renyi} H. Moradpour et al. [arXiv:1803.02195].
\bibitem{Komatsu} N. Komatsu, Eur. Phys. J. C {\bf77}, 229 (2017);\\
 H. Moradpour, A. Bonilla, E. M. C. Abreu, J. A. Neto,
Phys. Rev. D {\bf96}, 123504 (2017);\\ H. Moradpour, A. Sheykhi, C. Corda, I. G. Salako, Under review in Phys. Lett. B;
\\ H. Moradpour, Int. Jour. Theor. Phys. {\bf55}, 4176 (2016);\\ E. M. C. Abreu, J. Ananias Neto, A. C. R. Mendes, W.
Oliveira, Physica. A {\bf392}, 5154 (2013);\\ E. M. C. Abreu, J. Ananias Neto. Phys. Lett. B {\bf727},
524 (2013);\\ E. M. Barboza Jr., R. C. Nunes, E. M. C. Abreu, J. A.
Neto, Physica A: Statistical Mechanics and its Applica-
tions, {\bf436}, 301 (2015);\\ R. C. Nunes, et al. JCAP, {\bf08}, 051 (2016);\\ N. Komatsu, S. Kimura. Phys. Rev. D {\bf88}, 083534
(2013);\\ N. Komatsu, S. Kimura. Phys. Rev. D {\bf89}, 123501
(2014);\\ N. Komatsu, S. Kimura. Phys. Rev. D {\bf90}, 123516
(2014);\\ N. Komatsu, S. Kimura. Phys. Rev. D {\bf93}, 043530
(2016).
\bibitem{Tsallis} C. Tsallis, L. J. L. Cirto, Eur. Phys. J. C {\bf73} (2013) 2487.
\bibitem{bam} N. Saridakis, K. Bamba, R. Myrzakulov, [arXiv:1806.01301].
\bibitem{gdo1} G. Olivares, F. Atrio, D. Pavon, Phys. Rev. D {\bf71}, 063523 (2005).
\bibitem{gdo2} O. Bertolami , F. Gil Pedro, M. Le Delliou, Phys. Lett. B {\bf654}, 165 (2007).
\bibitem{ob1} A. A. Costa, X. D. Xu, B. Wang, E. G. M. Ferreira, and E. Abdalla, Phys. Rev. D {\bf89}, 103531 (2014).
\bibitem{ob2} X. D. Xu, B. Wang, and E. Abdalla, Phys. Rev. D {\bf 85}, 083513 (2012).
\bibitem{ob3} J. H. He, B. Wang, and E. Abdalla, Phys. Rev. D {\bf83}, 063515 (2011).
\bibitem{ob4} S. Wang, Y. Z. Wang, J. J. Geng, and X. Zhang, Eur. Phys. J. C {\bf74}, 3148 (2014).
\bibitem{ob5} J. H. He, B. Wang, E. Abdallab, and D. Pav\'{o}n, JCAP, {\bf12}, 022 (2010).
\bibitem{ob6} E. Abdalla, L. R. Abramo, and J. C. C. de Souza, Phy. Rev. D {\bf82}, 023508 (2010).
\bibitem{ob7} X. D. Xu, B. Wang, P. Zhang, and F. A. Barandela, JCAP, {\bf12}, 001 (2013).
\bibitem{pav} D. Pavon and W. Zimdahl, Phys. Lett. B {\bf628}, 206 (2005).
\bibitem{h1} M. Honarvaryan, A. Sheykhi and H. Moradpour, Int. J. Mod. Phys. D  {\bf24}, 1550048 (2015).
\bibitem{co1} L. Amendola, Phys. Rev. D {\bf62}, 043511 (2000).
\bibitem{co2} L. Amendola and C. Quercellini, Phys. Rev. D {\bf 68}, 023514 (2003).
\bibitem{co3} L. Amendola, S. Tsujikawa, and M. Sami, Phys. Lett. B {\bf632}, 155 (2006).
\bibitem{co4} S. del Campo, R. Herrera, and D. Pav\'{o}n, Phys. Rev. D {\bf78}, 021302 (2008).
\bibitem{co5} C. G. Bohmer, G. Caldera-Cabral, R. Lazkoz, and R. Maartens, Phys. Rev. D {\bf78}, 023505 (2008).
\bibitem{co6} S. Chen, B. Wang, and J. Jing, Phys. Rev. D {\bf78}, 123503 (2008).
\bibitem{Daly} R. A. Daly et al., Astrophys. J. {\bf 677} 1 (2008).
\bibitem{Gao} C. J. Gao, X. L. Chen and Y. G. Shen, Phys. Rev. D {\bf79}, 043511 (2009).
\end{thebibliography}
\end{document}